\documentclass{vldb}
\usepackage{graphicx}
\usepackage{balance}  
\usepackage{subfig}
\usepackage{hyperref}
\usepackage{tablefootnote}
\usepackage{xspace}
\usepackage{pgfplots}
\usepackage{filecontents}
\usepackage{blindtext}

\usepackage[bottom]{footmisc}

\definecolor{xblu}{RGB}{70,135,248}
\definecolor{xred}{RGB}{230,76,49}
\definecolor{xyellow}{RGB}{249,187,0}
\definecolor{xgreen}{RGB}{62,167,76}

\definecolor{xgrey}{RGB}{158,158,158}
\definecolor{xbrown}{RGB}{121,85,72}

\definecolor{xorange}{RGB}{243, 123, 29}
\definecolor{xpurple}{RGB}{128, 88, 165}
\definecolor{xblack}{RGB}{0, 0, 0}
\definecolor{xdarkblu}{RGB}{51, 102, 153}

\begin{document}

\title{An Experimental Survey on Big Data Frameworks}



%
%
%
%

\numberofauthors{5} 

\author{
%
%
\alignauthor
Wissem Inoubli\\
       \affaddr{University of Tunis El Manar, Faculty of Sciences of Tunis, LIPAH}\\
       \affaddr{Tunis, Tunisia}\\
       \email{inoubli.wissem@gmail.com}
\alignauthor 
Sabeur Aridhi\\
       \affaddr{University of Lorraine, CNRS, Inria, LORIA}\\
       \affaddr{F-54000 Nancy, France}\\
       \email{sabeur.aridhi@loria.fr}
\alignauthor 
Haithem Mezni\\
       \affaddr{University of Jendouba, SMART Lab}\\
       \affaddr{Avenue de l'Union du Maghreb Arabe, Jendouba 8189, Tunisia}\\
       \email{haithem.mezni@fsjegj.rnu.tn}
\and  
\alignauthor 
Mondher Maddouri\\
       \affaddr{College Of Buisness, University of Jeddah}\\
       \affaddr{P.O.Box 80327, Jeddah 21589 Kingdom of Saudi Arabia}\\
       \email{maddourimondher@yahoo.fr}
\alignauthor 
Engelbert Mephu Nguifo\\
       \affaddr{University of Clermont Auvergne, LIMOS}\\
       \affaddr{BP 10448, F-63000}\\
       \affaddr{Clermont-Ferrand, France}\\
       \email{mephu@isima.fr}
}

\maketitle

\begin{abstract}
Recently, increasingly large amounts of data are generated from a variety of sources.  
Existing data processing technologies are not suitable to cope with the huge amounts of generated data. 
Yet, many research works focus on Big Data, a \textit{buzzword} referring to the processing of massive volumes of (unstructured) data. 
Recently proposed frameworks for Big Data applications help to store, analyze and process the data. 
In this paper, we discuss the challenges of Big Data and we survey existing Big Data frameworks. We also present an experimental evaluation and a comparative study of the most popular Big Data frameworks with several representative batch and iterative workloads.  
This survey is concluded with a presentation of best practices related to the use of studied frameworks in several application domains such as machine learning, graph processing and real-world applications. 
\end{abstract}


\section{Introduction}
\label{sec:intro}
In recent decades, increasingly large amounts of data are generated from a variety of sources. 
The size of generated data per day on the Internet has already exceeded two exabytes~\cite{Gandomi2015137}. 
Within one minute, 72 hours of videos are uploaded to Youtube, around 30.000 new posts are created on the Tumblr blog platform, more than 100.000 Tweets are shared on Twitter and more than 200.000 pictures are posted on Facebook \cite{Gandomi2015137}. 

Big Data problems lead to several research questions such as (1) how to design scalable environments, (2) how to provide fault tolerance and (3) how to design efficient solutions. 
Most existing tools for storage, processing and analysis of data are inadequate for massive volumes of heterogeneous data. 
Consequently, there is an urgent need for more advanced and adequate Big Data solutions. 

Many definitions of Big Data have been proposed throughout the literature. Most of them agreed that Big Data problems share four main characteristics, referred to as the four V's (Volume, Variety, Veracity and Velocity) \cite{oguntimilehin}. The volume refers to the size of available datasets which typically require distributed storage and processing. The variety refers to the fact that Big Data is composed of several different types of data such as text, sound, image and video. The veracity refers to the biases, noise and abnormality in data. The velocity deals with the place at which data flows in from various sources like social networks, mobile devices and Internet of Things (IoT). 

In this paper, we first give an overview of most popular and widely used Big Data frameworks which are designed to cope with the above mentioned Big Data problems. We identify some key features which characterize Big Data frameworks. 
These key features include the programming model 
and the capability to allow for iterative processing of (streaming) data. 
We also give a categorization of existing frameworks according to the presented key features. Then, we present an experimental study on Big Data processing systems with several representative batch, stream and iterative workloads. 

Extensive surveys have been conducted to discuss Big Data Frameworks \cite{ref111} \cite{ref311} \cite{ref211}. 
However, our experimental survey differs from existing ones by the fact that it considers performance evaluation of popular Big Data frameworks from different aspects. 
In our work, we compare the studied frameworks in the case of both batch processing and stream processing which is not studied in existing surveys. 
We also mention that our experimental study is concluded by some best practices related to the usage of the studied frameworks in several application domains.  

More specifically, the contributions of this paper are the following:
\begin{itemize}
 \item We present an overview of most popular Big Data frameworks and we categorize them according to some features. 
 \item \textcolor{black}{We experimentally evaluate the performance of the presented frameworks and we present a comparative study of them in the case of both batch processing, stream processing. }
 \item We highlight best practices related to the use of popular Big Data frameworks in several application domains.  
\end{itemize}

The remainder of the paper is organized as follows. 
In Section \ref{sec:related}, we present existing surveys on Big Data frameworks and we highlight the motivation of our work. 
In Section  \ref{sec:cat}, we discuss existing Big Data frameworks and provide a categorization of them. 
In Section \ref{sec:experim}, we present a comparative study of the presented Big Data frameworks and we discuss the obtained results. 
In Section \ref{sec:best}, we present some best practices of the studied frameworks.  
Some concluding points are given in Section \ref{sec:conc}. 
\section{Related works}
 \label{sec:related}
 In this section, we highlight the existing surveys on Big Data frameworks and we describe their main contributions. From the ten discussed surveys, only six have experimentally studied some of the Big Data frameworks.


In \cite{rw1}, the authors compared several MapReduce implementations like Hadoop \cite{li2015mapreduce}, Twister \cite{twister} and LEMO-MR \cite{limomr} on many workloads.  Particularly, performance and scalability of the studied frameworks have been evaluated. 

In \cite{rw2}, an experimental study on Spark, Hadoop and Flink has been conducted. Mainly, the impact of some configuration parameters of the studied frameworks (e.g., number of mappers and reducers in Hadoop, number of threads in the case of Spark and Flink) on the runtime while running several workloads was studied. 

In \cite{mrvsspark}, the authors conducted an experimental study on Spark and Hadoop. They developed two profiling tools: (1) a study of the resource utilization for both MapReduce and Spark; (2) a break-down of the task execution time for in-depth analysis. The conducted experiments showed that Spark is about 2.5x, 5x, and 5x faster than MapReduce, for WordCount, k-means, and PageRank workloads, respectively. 

Some other works like \cite{ref111} \cite{rw5} \cite{rw6} \cite{ref211} tried to highlight Big Data fundamentals. They discussed the challenges related to Big Data applications and they presented the main features of some Big Data processing frameworks. 

Two works have compared Spark and Flink from theoretical and/or experimental point of view \cite{rw3} \cite{rw9}. Scalability and impact of the size on disk, as well as the performance of specific functionalities of the compared frameworks have been considered. In \cite{rw3}, the authors discussed the main difference between Spark and Flink and presented an empirical study of both frameworks in the case of machine learning applications. In \cite{rw9}, Marcu et al studied the impact of different architectural choices and parameter configurations on the perceived performance in the case of batch processing is studied. The performance of the studied frameworks has been evaluated with several representative batch and iterative workloads. 

The work presented in \cite{rw7} deals with in-memory Big Data management and processing frameworks. The authors provided a review of several in-memory data management and processing proposals and systems, including both data storage systems and data processing frameworks. They also presented some key factors that need to be considered in order to achieve efficient in-memory data management and processing, such as RDD for in-memory data persistence, immutable objects to improve response time, and data placement optimization. 




In \cite{rw11}, the authors conducted an experimental study on Storm and Flink in a stream processing context. The aim of the conducted study is to understand how current design aspects of modern stream processing systems interact with modern processors when running different types of applications. However, the study mainly focuses on evaluating the  common  design  aspects  of  stream processing systems on  scale-up  architectures,  rather  than  comparing  the  performance  of
individual systems. 

We mention that most of the above presented surveys are limited in terms of both the evaluated features of Big Data frameworks and  the number of considered frameworks. For example, in \cite{rw11}, only stream processing frameworks are considered while in \cite{rw1} \cite{rw2} \cite{rw3} \cite{rw9}, only batch processing frameworks are considered. We highlight that our experimental survey differs from the above presented works by the fact that it compares the studied frameworks in the case of both batch and stream processing. It also deals with several representative batch and iterative workloads which is not considered in most existing surveys. Add to that, additional parameters (e.g., memory, threads) are configured to better evaluate the discussed frameworks. Moreover, monitoring capacities differentiate our work from the existing surveys. In fact, a personalized tool is implemented for the different tests to effectively monitor resource usage.

\section{Big Data Frameworks}
\label{sec:cat}
In this section, we survey some popular Big Data frameworks and categorize them according to their key features. 
\textcolor{black}{ 
These key features are (1) the programming model, (2) the supported programming languages, (3) the type of data sources and 
(4) the capability to allow for iterative data processing, (5) the compatibility of the framework with existing machine learning libraries, and (6) the fault tolerance strategy.}

\textcolor{black}{
\textbf{Running Example.} Throughout this paper, we use the WordCount program as a running example in order to explain the studied frameworks. The WordCount example consists on reading a set of text files and counting how often words occur.
Snapshots of the codes used to implement the WordCount example with the studied frameworks  are available in this link \href{https://members.loria.fr/SAridhi/files/software/bigdata/}{https://members.loria.fr/SAridhi/files/software/bigdata/}.
}
\subsection{Apache Hadoop}
\subsubsection{Hadoop system overview}
Hadoop is an Apache project founded in 2008 by Doug Cutting at Yahoo and Mike Cafarella at the University of Michigan \cite{8polato2014comprehensive}.
Hadoop consists of two main components: (1) Hadoop Distributed File System (HDFS) for data storage and (2) Hadoop MapReduce, an implementation of the MapReduce programming model \cite{mapreduce}. 
In what follows, we discuss the MapReduce programming model,  HDFS and Hadoop MapReduce. 

\begin{figure}[t]
\centering
\includegraphics[width=0.5\textwidth]{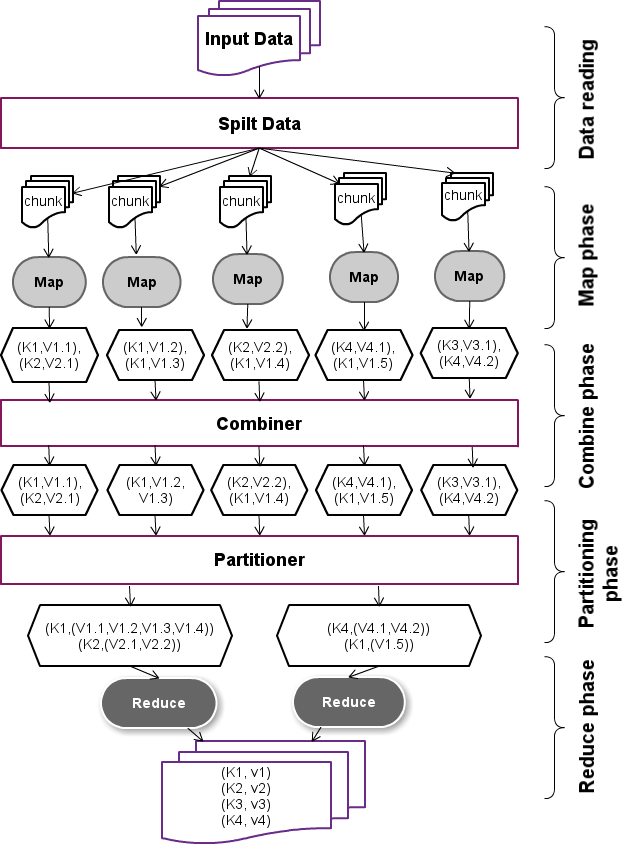}
\caption{The MapReduce architecture}
\label{fig:mr}
\end{figure}

\textbf{MapReduce Programming Model.}
MapReduce is a programming model that was designed to deal with parallel processing of large datasets. 
MapReduce has been proposed by Google in 2004 \cite{mapreduce} as an abstraction that allows to perform simple computations while hiding the details of parallelization, distributed storage, load balancing and enabling fault tolerance.
The central features of the MapReduce programming model are two functions, written by a user: Map and Reduce. 
The Map function takes a single key-value pair as input and produces a list of intermediate key-value pairs. 
The intermediate values associated with the same intermediate key are grouped together and passed to the Reduce function. 
The Reduce function takes as input an intermediate key and a set of values for that key. It merges these values together to form a smaller set of values. 
The system overview of MapReduce is illustrated in Fig.~\ref{fig:mr}. 

\begin{figure}[t]
\centering
\includegraphics[width=0.5\textwidth]{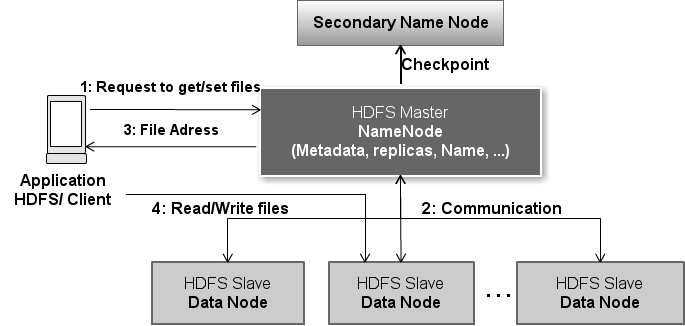}
\caption{HDFS architecture}
\label{fig:Hadoop}
\end{figure}

As shown in Fig.~\ref{fig:mr}, the basic steps of a MapReduce program are as follows: 
\begin{enumerate}
\item \textbf{Data reading:} in this phase, the input data is transformed to a set of key-value pairs. The input data may come from various sources such as \textcolor{black}{file systems, database management systems or main memory (RAM)}. The input data is split into several fixed-size chunks. Each chunk is processed by one instance of the Map function.
\item \textbf{Map phase:} for each chunk having the key-value structure, the corresponding Map function is triggered and produces a set of intermediate key-value pairs. 
\item \textbf{Combine phase:} this step aims to group together all intermediate key-value pairs associated with the same intermediate key.
\item \textbf{Partitioning phase:} following their combination, the results are distributed across the different Reduce functions.
\item \textbf{Reduce phase:} the Reduce function merges key-value pairs having the same key and computes a final result. 
\end{enumerate}
\textbf{HDFS.}
HDFS is an open source implementation of the distributed Google File System (GFS) \cite{gfs}. 
It provides a scalable distributed file system for storing large files over distributed machines in a reliable and efficient way \cite{9white2012hadoop}. 
In Fig.~\ref{fig:Hadoop}, we show the abstract architecture of HDFS and its components. 
It consists of a master/slave architecture with a Name Node being master and several Data Nodes as slaves. 
The Name Node is responsible for allocating physical space to store large files sent by the HDFS client. 
If the client wants to retrieve data from HDFS, it sends a request to the Name Node. 
The Name Node will seek their location in its indexing system and subsequently sends their address back to the client. 
The Name Node returns to the HDFS client the meta data (filename, file location, etc.) related to the stored files. 
A secondary Name Node periodically saves the state of the Name Node. If the Name Node fails, the secondary Name Node takes over automatically.

\begin{figure}[t]
\centering
\includegraphics[width=0.5\textwidth]{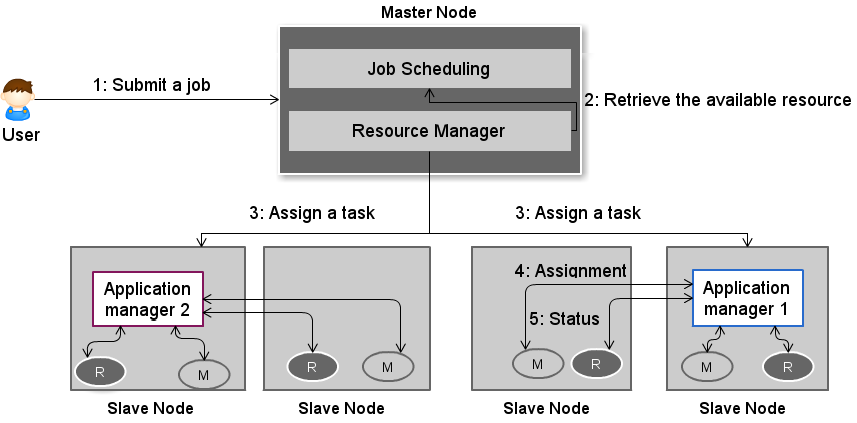}
\caption{YARN architecture}
\label{fig:YARN}
\end{figure}

\textbf{Hadoop MapReduce.}
There are two main versions of Hadoop MapReduce. 
In the first version called MRv1, Hadoop MapReduce is essentially based on two components: 
(1) the Task Tracker that aims to supervise the execution of the Map/Reduce functions and 
(2) the Job Tracker which represents the master part and allows resource management and job scheduling/monitoring. The Job Tracker supervises and manages the Task Trackers \cite{9white2012hadoop}.
In the second version of Hadoop called YARN, the two major features of the Job Tracker have been split into separate daemons: 
(1) a global Resource Manager and (2) per-application Application Master.  
In Fig.~\ref{fig:YARN}, we illustrate the overall architecture of YARN. 

As shown in Fig.~\ref{fig:YARN}, the Resource Manager receives and runs MapReduce jobs. 
The per-application Application Master obtains resources from the ResourceManager and works with the Node Manager(s) to execute and monitor the tasks. 
In YARN, the Resource Manager (respectively the Node Manager) replaces the Job Tracker (respectively the Task Tracker) \cite{li2015mapreduce}.\\
\textcolor{black}{
Note that other well-know cluster managers are heavily used by Big Data systems. Taking as examples Mesos \cite{mesos} and Zookeeper \cite{zk}.\\
Mesos is an open source cluster manager that ensures a dynamic resources sharing and provides efficient resources management for distributed frameworks \cite{mesos}. It is based on a master/slave architecture. The master node relies on a daemon, called \textit{master process}. This later manages all \textit{executor} daemons deployed in the slave nodes, on which user tasks are distributed and executed.\\
 Apache ZooKeeper is an open source and fault-tolerant coordinator for large distributed systems \cite{zk}. It provides a centralized service for maintaining the cluster\textsc{\char13}s configuration and management. It also ensures the data or service synchronization in distributed applications. Unlike YARN or Mesos, Zookeeper is based on a cooperative control architecture, where the same service is deployed in all machines of the cluster. Each client or application can request the Zookeeper service by connecting to any machine in the cluster.
}
\subsubsection{WordCount example with Hadoop  }
\textcolor{black}{
A WordCount program in Hadoop consists of a MapReduce job that counts the number of occurrences of each word in a file stored in the HDFS.  The Map task maps the text data in the file and counts each word in the data chunk provided to the Map function (see Fig.~\ref{fig:mr}). The result of the Map tasks are passed to Reduce function which combines and reduces the data to generate the final result. }
\subsection{Apache Spark}
\subsubsection{Spark system overview}

Apache Spark is a powerful processing framework that provides an ease of use tool for efficient analytics of heterogeneous data. 
It was originally developed at UC Berkeley in 2009 \cite{Zaharia:2010:SCC:1863103.1863113}. 
Spark has several advantages compared to other Big Data frameworks like Hadoop and storm. 
Spark is used by many companies such as Yahoo, Baidu, and Tencent. 
A key concept of Spark is Resilient Distributed Datasets (RDDs). An RDD is basically an immutable collection of objects spread across a Spark cluster. 
In Spark, there are two types of operations on RDDs: (1) transformations and (2) actions.
Transformations consist in the creation of new RDDs from existing ones using functions like map, filter, union and join. 
Actions consist of final result of RDD computations.

 \begin{figure}[t]
\centering
\includegraphics[width=0.5\textwidth]{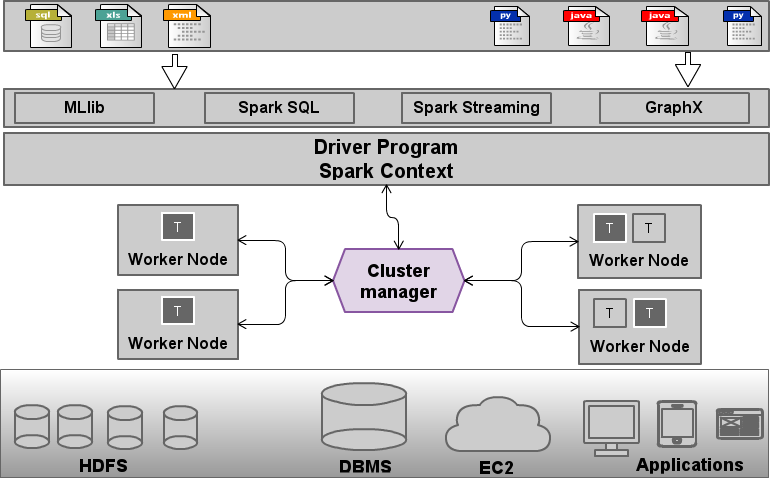}
\caption{Spark system overview}
\label{fig:Spark}
\end{figure} 
In Fig.~\ref{fig:Spark}, we present an overview of the Spark architecture. 
A Spark cluster is based on a master/slave architecture with three main components: 
\begin{itemize}
    \item \textbf{Driver Program:} this component represents the slave node in a Spark cluster. 
    It maintains an object called SparkContext that manages and supervises running applications.
    \item \textbf{Cluster Manager:} this component is responsible for orchestrating the workflow of application assigned by Driver Program to workers. 
    It also controls and supervises all resources in the cluster and returns their state to the Driver Program.  
    \item \textbf{Worker Nodes:} each Worker Node represents a container of one operation during the execution of a Spark program.  
\end{itemize}

Spark offers several Application Programming Interfaces (APIs) \cite{Zaharia:2010:SCC:1863103.1863113}:
\begin{itemize}
    \item \textbf{SparkCore:} 
Spark Core is the underlying general execution engine for the Spark platform. All other features and extensions are built on top of it. 
Spark Core provides in-memory computing capabilities and a generalized execution model to support a wide variety of applications, as well as Java, Scala, and Python APIs for ease of development.
    \item \textbf{SparkStreaming:} 
Spark Streaming enables powerful interactive and analytic applications across both streaming and historical data, while inheriting Spark's ease of use and fault tolerance characteristics. 
    It can be used with a wide variety of popular data sources including HDFS, Flume \cite{Chambers:2010:FEE:1806596.1806638}, Kafka \cite{Garg:2013:AK:2588385}, and Twitter \cite{Zaharia:2010:SCC:1863103.1863113}.
    \item \textbf{SparkSQL:} 
    Spark offers a range of features to structure data retrieved from several sources. 
    It allows subsequently to manipulate them using the SQL language \cite{Armbrust:2015:SSR:2723372.2742797}. 
    \item \textbf{SparkMLLib:} 
   Spark provides a scalable machine learning library that delivers both high-quality algorithms (e.g., multiple iterations to increase accuracy) and high speed (up to 100x faster than MapReduce) \cite{Zaharia:2010:SCC:1863103.1863113}. 
        \item \textbf{GraphX:} 
        GraphX \cite{Xin:2013:GRD:2484425.2484427} is a Spark API for graph-parallel computation (e.g., PageRank algorithm and collaborative filtering). 
        At a high-level, GraphX extends the Spark RDD abstraction by introducing the Resilient Distributed Property Graph: a directed multigraph with properties attached to each vertex and edge. 
        To support graph computation, GraphX provides a set of fundamental operators (e.g., subgraph, joinVertices, and MapReduceTriplets) as well as an optimized variant of the Pregel API \cite{15malewicz2010pregel}. 
        In addition, GraphX includes a growing collection of graph algorithms (e.g., PageRank, Connected components, Label propagation and Triangle count) to simplify graph analytics tasks. 
\end{itemize} 
\subsubsection{WordCount example with Spark}
\begin{figure}[t]
\centering
\includegraphics[width=0.5\textwidth]{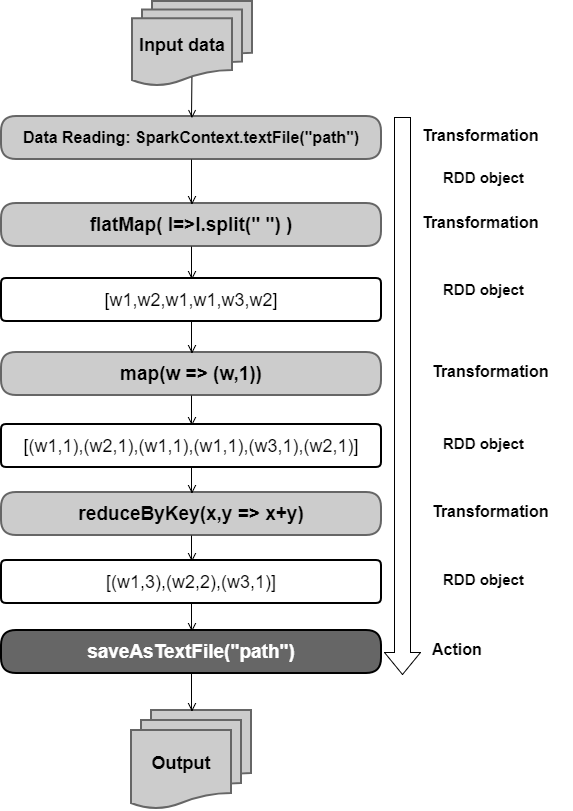}
\caption{\textcolor{black}{WordCount example with Spark }}
\label{fig:sparkwc}
\end{figure}

\textcolor{black}{In Spark, every job is modeled as a graph. The nodes of the graph represent transformations and/or actions, whereas the edges represent data exchange between the nodes through RDD objects. Through Fig.~\ref{fig:sparkwc}, we show the execution plan for a WordCount job. In the first step, the SparkContext object is used to read the input data from any sources (e.g., HDFS) and to create an RDD. In the second step, several operations can be applied to the RDD. In this example, we apply a \textit{flatMap} operation that receives the lines of RDD, and applies a lambda function to each line of the RDD in order to generate a set of words. Then, a \textit{map} function is applied in order to create a set of key-value pairs, in which the key is a word and the value is the number one. The next step consists on computing the sum of the values of each key using the \textit{reduceByKey} function.
The final results are written using the \textit{saveAsFile} function.}

\subsection{Apache Storm}
\subsubsection{Storm system overview}
Storm \cite{Toshniwal:2014:STO:2588555.2595641}  is an open source framework for processing large structured and unstructured data in real time. 
storm is a fault tolerant framework that is suitable for real time data analysis, machine learning, sequential and iterative computation.
Following a comparative study of storm and Hadoop, we find that the first is geared for real time applications while the second is effective for batch applications. 

\begin{figure}[t]
\centering
\includegraphics[width=0.5\textwidth]{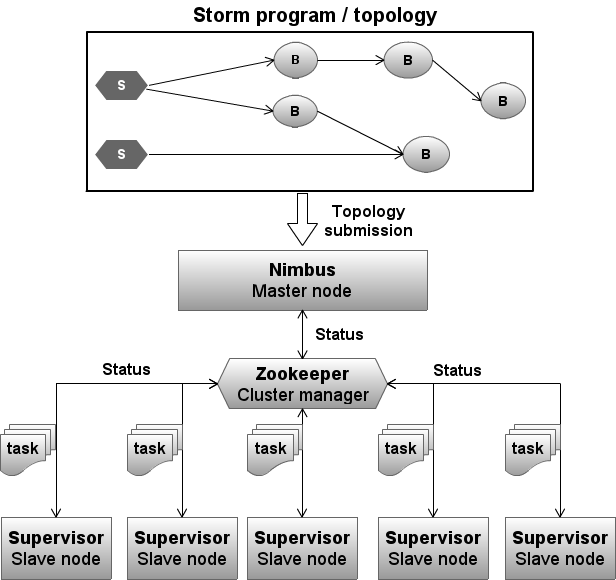}
\caption{Topology of a Storm program and architecture}
\label{fig:top}
\end{figure}

As shown in Fig.~\ref{fig:top}, a storm program/topology is represented by a directed acyclic graphs (DAG). The edges of the program DAG represent data transfer. 
The nodes of the DAG are divided into two types: spouts and bolts. 
The spouts (or entry points) of a storm program represent the data sources. 
The bolts represent the functions to be performed on the data. Note that storm distributes bolts across multiple nodes to process the data in parallel. 

In Fig.~\ref{fig:top}, we show a storm cluster administrated by zookeeper, a service for coordinating processes of distributed applications \cite{Hunt:2010:ZWC:1855840.1855851}. storm is based on two daemons called Nimbus (in master node) and supervisor (for each slave node).  
Nimbus supervises the slave nodes and  assigns tasks to them. 
If it detects a node failure in the cluster, it re-assigns the task to another node. Each supervisor controls the execution of its tasks (affected by the nimbus). 
It can stop or start the spots following the instructions of Nimbus. 
Each topology submitted to storm cluster is divided into several tasks. 
\subsubsection{WordCount example with Storm}
\begin{figure}[t]
\centering
\includegraphics[width=0.5\textwidth]{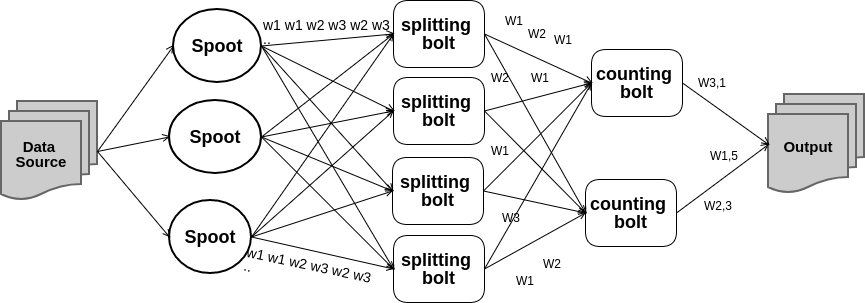}
\caption{\textcolor{black}{WordCount example with Storm}}
\label{fig:stormwc}
\end{figure}

\textcolor{black}{
Since Storm is a framework for stream processing, we run the WordCount example in stream mode. A Storm WordCount job consists on a topology that combines a set of spoots and bolts, where the spoots are used to get the data and the bolts are used to process the data. 
In Fig.~\ref{fig:stormwc}, three processing layers are used to process the data.  In the first layer, the spoots are used to read the input data from the sources and push the data (as lines of text) to the next layer. Then, in the next layer, a set of bolts are used to generate a set of words from each consumed line (from the previous layer). Finally, the last bolts are used to count for each word its number of occurrences.}
\subsection{Apache Samza}
\subsubsection{Samza system overview}
\begin{figure}[t]
\centering
\includegraphics[width=0.5\textwidth]{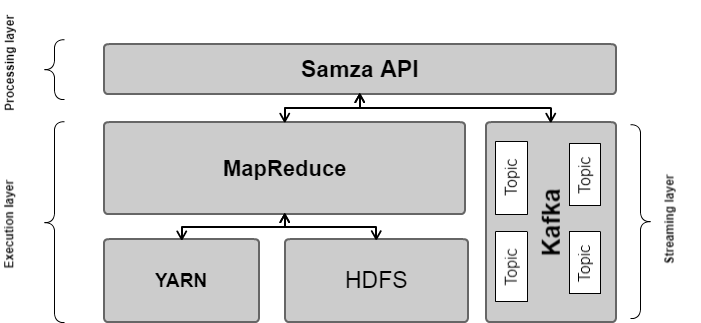}
\caption{\textcolor{black}{Samza architecture}}
\label{fig:smaza}
\end{figure}

\textcolor{black}{Apache Samza \cite{samza} is a distributed processing framework created by LinkedIn to solve various  kinds of stream processing requirements such as tracking data, service logging, and data ingestion pipelines for real-time services. Since then, it was adopted and deployed in several projects. Samza is designed to handle large messages and to provide file system persistence for them. It uses Apache Kafka as a distributed broker for messaging, and YARN for distributed resource allocation and scheduling. YARN resource manager is adopted by Samza to provide fault tolerance, processor isolation, security, and resource management in the used cluster.
As illustrated in Fig.~\ref{fig:smaza}, Samza is based on three layers. The first layer is devoted to streaming data and uses Apache Kafka to manage the data flow. The second layer is based on YARN resource manager to handle the distributed execution of Samza jobs and to manage CPU and memory usage across a multi-tenant cluster of machines. The processing capabilities are available in the third layer which represents the Samza core and provides APIs for creating and running stream tasks in  cluster \cite{samza}. In this layer, several abstract classes can be implemented by the user to perform specific processing tasks. 
These abstract classes could be implemented with MapReduce, in order to ensure the distributed processing.
}

\subsubsection{WordCount example with Samza}
\begin{figure}[t]
\centering
\includegraphics[width=0.5\textwidth]{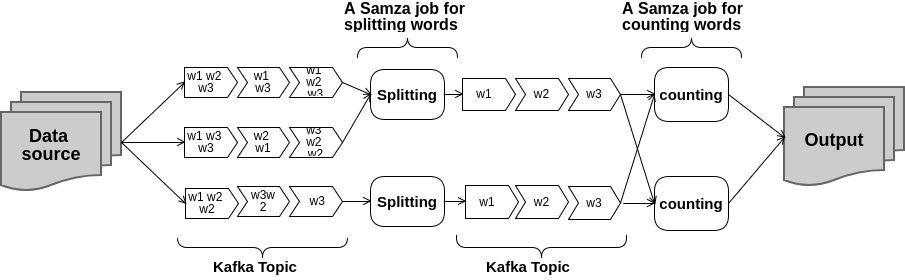}
\caption{\textcolor{black}{WordCount example with Samza}}
\label{fig:samzamwc}
\end{figure}

\textcolor{black}{
A Samza job is usally based on tow parts. The first part is responsible for data processing and the second part is responsible for data flow transfer between the data processing units. As shown in Fig~\ref{fig:samzamwc}. shows the execution steps of a wordCount job with Samza. In the first step, the data is read from the source and sent to the first Samza  task, called splitter, through a kafka topic. In this step, each message is splitted into a set of words.
In the next step, another Samza task called counter consumes the set of words, and counts for each one the number of occurrences and generates the final result.
}
\subsection{Apache Flink}
\subsubsection{Flink system overview}

\begin{figure}[t]
\centering
\includegraphics[width=0.5\textwidth]{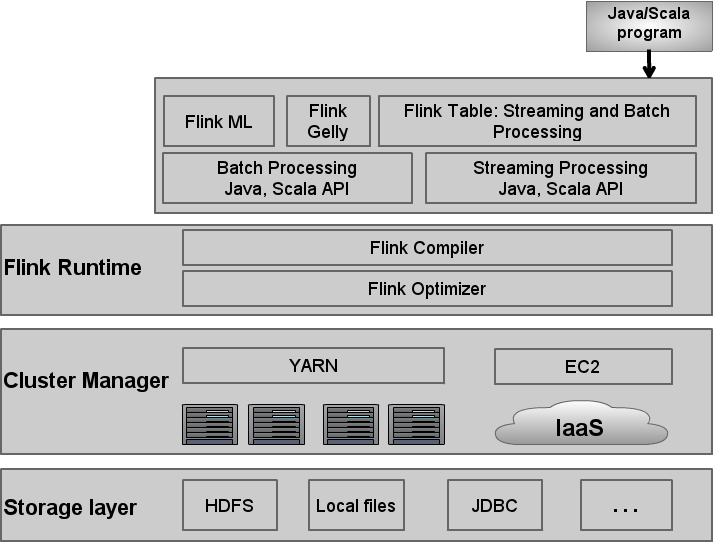}
\caption{Flink architecture}
\label{fig:Flink}
\end{figure}

Flink \cite{Alexandrov:2014:SPB:2691523.2691544} is an open source framework for processing data in both real time mode and batch mode.
It provides several benefits such as fault-tolerant and large scale computation. 
The programming model of Flink is similar to MapReduce. 
By contrast to MapReduce, Flink offers additional high level functions such as join, filter and aggregation.
Flink allows iterative processing and real time computation on stream data collected by different tools such as Flume \cite{Chambers:2010:FEE:1806596.1806638} and Kafka \cite{Garg:2013:AK:2588385}. 
It offers several APIs on a more abstract level allowing the user to launch distributed computation in a transparent and easy way. 
Flink ML is a machine learning library that provides a wide range of learning algorithms to create fast and scalable Big Data applications.
In Fig.~\ref{fig:Flink}, we illustrate the architecture and components of Flink. 

As shown in Fig.~\ref{fig:Flink}, the Flink system consists of several layers. 
In the highest layer, users can submit their programs written in Java or Scala. User programs are then converted by the Flink compiler to DAGs. 
Each submitted job is represented by a graph. Nodes of the graph represent operations (e.g., map, reduce, join or filter) that will be applied to process the data. 
Edges of the graph represent the flow of data between the operations.
A DAG produced by the Flink compiler is received by the Flink optimizer in order to improve performance by optimizing the DAG (e.g., re-ordering of the operations). 
The second layer of Flink is the cluster manager which is responsible for planning tasks, monitoring the status of jobs and resource management.
The lowest layer is the storage layer that ensures storage of the data to multiple destinations such as HDFS and local files.

\subsubsection{WordCount example with Flink }
\begin{figure}[t]
\centering
\includegraphics[width=0.5\textwidth]{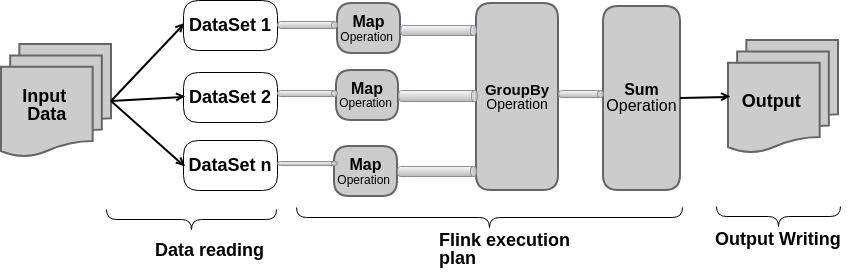}
\caption{\textcolor{black}{WordCount with Flink}}
\label{fig:flinkwc}
\end{figure}

\textcolor{black}{
In order to implement the WordCount example with Flink, we can use the abstract functions provided by Flink such as \textit{map}, \textit{flatMap} and \textit{groupBy}. 
First, the input data is read from the data source and stored in several dataset objects. Then, a \textit{map} operation is applied to the dataset objects in order to generate key-value pairs, with the word as a key and one as value. Then, the \textit{groupBy} function is applied to aggregate the list of key-value pairs generated in the previous step (see Fig.~\ref{fig:flinkwc}). 
Finally, the number of occurrences of each word is calculated using the \textit{sum} function and the final  results are generated.
}
\subsection{Categorization of Big Data Frameworks}
\begin{table*}[ht]
\centering
 \scalebox{0.8}{
\begin{tabular}{p{3.5cm}p{2.5cm}p{2.5cm}p{2.5cm}p{2.5cm} p{2.5cm}}
  \cline{2-6}
    \textbf{}  & \textbf{Hadoop} & \textbf{Spark} & \textbf{Storm} & \textbf{Flink} & \textbf{\textcolor{black}{Samza}} \\
   \hline
    \textbf{Data format}&Key-value & \textcolor{black}{Key-value, RDD} & Key-value & Key-value & \textcolor{black}{Events} \\
   \hline
    \textbf{Processing mode} & Batch & Batch and Stream & Stream & Batch and Stream & \textcolor{black}{Stream} \\
   \hline
    \textbf{Data sources} & HDFS & HDFS, DBMS and Kafka & HDFS, HBase and Kafka & Kafka, Kinesis, message queus, socket streams and files & \textcolor{black}{Kafka} \\
   \hline
      \textbf{Programming model} & Map and Reduce & Transformation and Action & Topology & Transformation &  \textcolor{black}{Map and Reduce}\\
   \hline
   \textbf{Supported programming language} & Java & Java, Scala and Python & Java & Java & \textcolor{black}{Java} \\
   \hline
   \textbf{Cluster manager} & YARN & Standalone, YARN and Mesos & YARN ~or~Zookeeper & Zookeeper & \textcolor{black}{YARN} \\
   \hline
   \textbf{Comments} & Stores large data in HDFS  & Gives several APIs to develop interactive applications  & Suitable for real-time applications & Flink is an extension of MapReduce with graph methods & \textcolor{black}{Based on Hadoop and Kafka} \\
   \hline
    \textbf{Iterative computation} & Yes (by running multiple MapReduce jobs) & Yes & Yes & Yes & \textcolor{black}{YES} \\
   \hline
   
      \textbf{\textcolor{black}{Interactive Mode}} & \textcolor{black}{No}  & \textcolor{black}{Yes} & \textcolor{black}{No}  & \textcolor{black}{No} & \textcolor{black}{No} \\
   \hline
\textbf{\textcolor{black}{Machine learning compatibility}} & \textcolor{black}{Mahout}  & \textcolor{black}{SparkMLlib} &  \textcolor{black}{Compatible with SAMOA API}   & \textcolor{black}{FlinkML} &\textcolor{black}{ Compatible with SAMOA API} \\
   \hline
   
   \textbf{\textcolor{black}{Fault tolerance}} & \textcolor{black}{Duplication feature}   & \textcolor{black}{Recovery technique on the RDD objects} &  \textcolor{black}{Checkpoints}   & \textcolor{black}{Checkpoints} & \textcolor{black}{Data partitioning} \\
   \hline
 \end{tabular} 
 }
 \caption{ A comparative study of popular Big Data frameworks}\label{tab:tab1}
 \end{table*}
\textcolor{black}{We present in Table~\ref{tab:tab1} a categorization of the presented frameworks according to data format, processing mode, used data sources, programming model, supported programming languages, cluster manager, machine learning compatibility, fault tolerance strategy and whether the framework allows iterative computation or not. }
%

\textcolor{black}{
As shown in Table~\ref{tab:tab1}, Hadoop, Flink and Storm use the key-value format to represent their data. This is motivated by the fact that the key-value format allows access to heterogeneous data. For Spark, both RDD and key-value models are used to  allow fast data access. 
We have also classified the studied big data frameworks into two categories: (1) batch mode and (2) stream mode. We have shown in Table~\ref{tab:tab1} that Hadoop processes the data  in batch mode, whereas the other frameworks allow the stream processing mode. 
 In terms of physical architecture, we notice tha all the studied frameworks are deployed in a cluster architecture,  and each framework uses a specified cluster manager. 
 We note that most of the studied frameworks use YARN as cluster manager. 
From a technical point of view, we mention that all the presented frameworks provide APIs for several programming languages like Java, Scala and Python. 
Each framework provides a set of abstract functions that are used to define the desired computation. 
We also presented in Table~\ref{tab:tab1} weather the studied framework provide a machine learning library or not. We notice that Spark and Flink provide their own machine learning libraries, while the other frameworks have some compatibility with other tools, such as SAMOA for Samza and Mahout for Hadoop.
}  

It is important to mention that Hadoop is currently one of the most widely used parallel processing solutions. Hadoop ecosystem consists of a set of tools such as Flume, HBase, Hive and Mahout. Hadoop is widely adopted in the management of large-size clusters. Its YARN daemon makes it a suitable choice to configure Big Data solutions on several nodes \cite{haste}. For instance, Hadoop is used by Yahoo to manage 24 thousands of nodes. Moreover, Hadoop MapReduce was proven to be the best choice to deal with text processing tasks \cite{Lin:2010:DTP:1855013}. 
We notice that Hadoop can run multiple MapReduce jobs to support iterative computing but it does not perform well because it can not cache intermediate data in memory for faster performance.

As shown in Table~\ref{tab:tab1}, Spark importance lies in its in-memory features and micro-batch processing capabilities, especially in iterative and incremental processing \cite{Bajaber2016}. 
In addition, Spark offers an interactive tool called SparkShell which allows to exploit the Spark cluster in real time. Once interactive applications were created, they may subsequently be executed interactively in the cluster. 
We notice that Spark is known to be very fast in some kinds of applications due to the concept of RDD and also to the DAG-based programming model. 

Flink shares similarities and characteristics with Spark. It offers good processing performance when dealing with complex Big Data structures such as graphs. Although there exist other solutions for large-scale graph processing, Flink and Spark are enriched with specific APIs and tools for machine learning, predictive analysis and graph stream analysis \cite{Alexandrov:2014:SPB:2691523.2691544} \cite{Zaharia:2010:SCC:1863103.1863113}.

\subsection{Real-world applications}
In this sub-section, we discuss the use of the studied frameworks in sevral real-world applications including health core applications, recommender systems, social network analysis and smart cites.
\subsubsection{Healthcare applications}
Healthcare scientific applications, such as body area network provide monitoring capabilities to decide on the health status of a host. This requires deploying hundreds of interconnected sensors over the human body to collect various data including breath, cardiovascular, insulin, blood, glucose and body temperature \cite{Zhang:2015:TAM:2943467.2943721}.
However, sending and processing iteratively such stream of health data is not supported by the original MapReduce model. Hadoop was initially designed to process big data already available in the distributed file system. In the literature, many extensions have been applied to the original Mapreduce model in order to allow iterative computing such as Haloop system \cite{haloop} and Twister \cite{twister}. Nevertheless, the two caching functionalities in Haloop that allow reusing processing data in the later iterations and make checking for a fix-point lack efficiency. Also, since processed data may partially remain unchanged through the different iterations, they have to be reloaded and reprocessed at each iteration. This may lead to resource wastage, especially network bandwidth and processor resources.
Unlike Haloop and existing MapReduce extensions, Spark provides support for interactive queries and iterative computing. RDD caching makes Spark efficient and performs well in iterative use cases that require multiple treatments on large in-memory datasets \cite{Bajaber2016}. 
\subsubsection{Recommendation systems}
Recommender systems is another field that began to attract more attention, especially with the continuous changes and the growing streams of users' ratings \cite{recsys}. Unlike traditional recommendation approaches that only deal with static item and user data, new emerging recommender systems must adapt to the high volume of item information and the big stream of user ratings and tastes. In this case, recommender systems must be able to process the big stream of data. For instance, news items are characterized by a high degree of change and user interests vary over time which requires a continuous adjustment of the recommender system. In this case, frameworks like Hadoop are not able to deal with the fast stream of data (e.g. user ratings and comments), which may affect the real evaluation of available items (e.g. product or news). In such a situation, the adoption of effective stream processing frameworks is encouraged in order to avoid overrating or incorporating user/item related data into the recommender system. Tools like Mahout, Flinkml and Sparkmllib include collaborative filtering algorithms, that may be used for e-commerce purpose and in some social network services to suggest suitable items to users \cite{clef}. 
\subsubsection{Social media}
Social media is another representative data source for big data that requires real-time processing and results. Its is generated from a wide range of Internet applications and Web sites including social and business-oriented networks (e.g. LinkedIn, Facebook), online mobile photo and video sharing services (e.g. Instagram, Youtube, Flickr), etc. 
This huge volume of social data requires a set of methods and algorithms related to, text analysis, information diffusion, information fusion, community detection and network analytics, which may be exploited to analyse and process information from social-based sources \cite{BelloOrgaz201645}. This also requires iterative processing and learning capabilities and necessitates the adoption of in-stream frameworks such as Storm and Flink along with their rich libraries.
\subsubsection{Smart cities}
Smart city is a broad concept that encompasses economy, governance, mobility, people, environment and living \cite{smart}. It refers to the use of information technology to enhance quality, performance and interactivity of urban services in a city. It also aims to connect several geographically distant cities \cite{booksmart}. Within a smart city, data is collected from sensors installed on utility poles, water lines, buses, trains and traffic lights. The networking of hardware equipment and sensors is referred to as Internet of Things (IoT) and represents a significant source of Big data. Big data technologies are used for several purposes in a smart city including traffic statistics, smart agriculture, healthcare, transport and many others \cite{booksmart}. 
For example, transporters of the logistic company UPS are equipped with operating sensors and GPS devices reporting the states of their engines and their positions respectively. This data is used to predict failures and track the positions of the vehicles. Urban traffic also provides large quantities of data that come from various sensors (e.g., GPSs, public transportation smart cards, weather conditions devices and traffic cameras). To understand this traffic behaviour, it is important to reveal hidden and valuable information from the big stream/storage of data. 
Finding the right programming model is still a challenge because of the diversity and the growing number of services \cite{Piro2014169}. Indeed, some use cases are often slow such as urban planning and traffic control issues. Thus, the adoption of a batch-oriented framework like Hadoop is sufficient. Processing urban data in micro-batch fashion is possible, for example, in case of eGovernment and public administration services. Other use cases like healthcare services (e.g. remote assistance of patients) need decision making and results within few milliseconds. In this case, real-time processing frameworks like Storm are encouraged. Combining the strengths of the above discussed frameworks may also be useful to deal with cross-domain smart ecosystems also called big services \cite{bigservice}.

\section{Experiments}
\label{sec:experim}
We have performed an extensive set of experiments to highlight the strengths and weaknesses of popular Big Data frameworks. The performed analysis covers scalability, impact of several configuration parameters on the performance and resource usage. For our tests, we evaluated Spark, Hadoop, Flink, Samza and Storm. For reproducibility reasons, we provide information about the implementation details and the used datasets in the following link:  \href{https://members.loria.fr/SAridhi/files/bigdata/}{https://members.loria.fr/SAridhi/files/bigdata/}. 
In this section, we describe the experimental setup and we discuss the obtained results. 
\subsection{Experimental environment}
\textcolor{black}{
All the experiments were performed in a cluster of 10 machines operating with Linux Ubuntu 16.04. Each machine is equipped with a 4 CPU, 8GB of main memory and 500 GB of local storage. For our tests, we used Hadoop 2.9.0, Flink 1.3.2, Spark 1.6.0, Samza 0.10.3 and Storm 1.1.1. All the studied frameworks have been deployed with YARN as a cluster manager. 
We also varied these parameters in order to analyze the impact of some of them on the performance of the studied frameworks. 
}
\subsection{Experimental protocol}
\label{sec:experimProtocol}
We consider two scenarios according to the data processing mode (Batch and Stream) of the evaluated frameworks.
\begin{itemize}
\item In the \textit{Batch mode} scenario, we evaluate Hadoop, Spark and Flink while running the WordCount, K-means and PageRank workloads with real and synthetic data sets.
In the WordCount application, we used tweets that are collected by Apache Flume \cite{Chambers:2010:FEE:1806596.1806638} and stored in HDFS. As shown in Fig\ref{fig:batch}, the collected data may come from different sources including social networks, local files, log files and sensors. In our case, Twitter is the main source of our collected data. The motivation behind using Apache Flume to collect the processed tweets is its integration facility in the Hadoop ecosystem (especially the HDFS system). Moreover, Apache Flume allows data collection in a distributed way and offers high data availability and fault tolerance. We collected 10 billions tweets and we used them to form large tweet files with a size on disk varying from 250 MB to 100 GB of data. 

\begin{figure*}[t]
\centering
\includegraphics[width=0.5\textwidth]{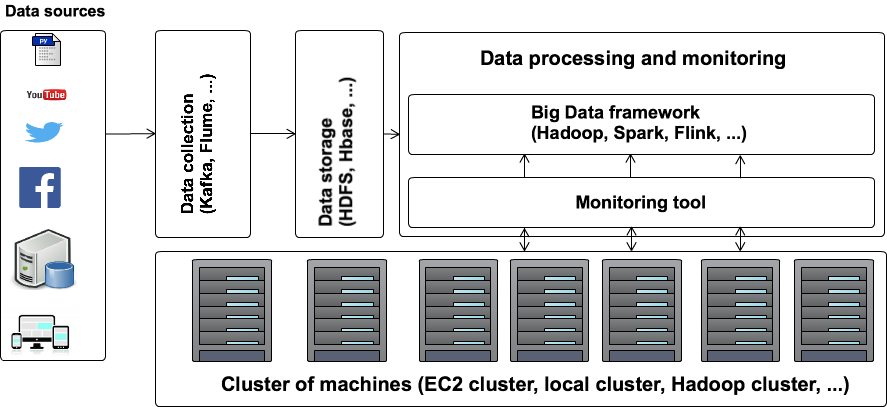}
\caption{\textit{Batch Mode} scenario}
\label{fig:batch}
\end{figure*}
\textcolor{black}{For K-means, we generated a synthetic data sets containing between 10000 and 100 millions learning examples.}
\textcolor{black}{For PageRank workload, we have used seven real graph datasets with different numbers of nodes and edges.} 
\textcolor{black}{Table \ref{tab:tab2} shows more details of the used datasets. \\
The above presented datasets have been downloaded from the Stanford Large Network Dataset Collection (SNAP) \footnote{\url{https://snap.stanford.edu/data/}} and formatted as plan files in which each line represents a link between two  nodes. 
We implemented the PageRank workload with Hadoop using a three-jobs workflow. In the first job, we read data from the text file and we generated a set of links for each page. The second job is responsible for setting an initial score for each page. The last job iteratively computes and sorts the pages' scores. Regarding the PageRank implementation with Spark, we followed the same execution logic as in Hadoop. We implemented a Spark job that applies the \textit{flatMap} function to generate key-value pairs for the corresponding links, and the \textit{map} function to initialize an initial score for each page. Finally, the \textit{reduceByKey} function is used to iteratively aggregate the page\textsc{\char13}s scores. As for the implemented Flink job, it starts by generating the page-score pairs using the \textit{flatMap} function. Then, it iteratively aggregates the scores for each page using the \textit{groupBy} function. Finally, it computes the total score of each page by applying the \textit{sum} function.
}

\begin{table*}
\centering
    \begin{tabular}{|c|l|l|l|}
   \textbf{ Dataset} & \textbf{Number of nodes} & \textbf{Number of edges} & \textbf{Description }            \\
    G1           & 685 230          & 7 600 595         & Web graph of Berkeley and Stanford \\
    G2           & 875 713          & 5 105 039         & Web graph from Google
    \\
  G3            & 325 729          & 1 497 134         & Web graph of Notre Dame            \\
    G4           & 281 903          & 2 312 497         & Web graph of Stanford              \\
    \textcolor{black}{G5}          & \textcolor{black}{1,965,206}         & \textcolor{black}{2,766,607}         & \textcolor{black}{RoadNet-CA }
     \\
      \textcolor{black}{G6}            & \textcolor{black}{3,997,962}         & \textcolor{black}{34,681,189}         & \textcolor{black}{Com-LiveJournal}
     \\
    \textcolor{black}{G7}            & \textcolor{black}{4,847,571}         & \textcolor{black}{68,993,773}        & \textcolor{black}{Soc-LiveJournal}
     \\
    \end{tabular}
    \caption{Graph datasets.}\label{tab:tab2}
\end{table*}

\item In the \textit{Stream mode} scenario, we evaluate real-time data processing capabilities of Storm, Flink, Samza and Spark. The Stream mode scenario is divided into three main steps. As shown in Fig. \ref{fig:stream}, the first step is devoted to data storage. To do this step, we collected 1 billion tweets from Twitter using Flume and we stored them in HDFS. The stored data is then transferred to Kafka, a messaging server that guarantees fault tolerance during the streaming and message persistence \cite{Garg:2013:AK:2588385}. The second step consists on sending the tweets as streams to the studied frameworks. To allow simultaneous streaming of the data collected from HDFS by Storm, Spark, Samza and Flink, we have implemented a script that accesses the HDFS and transfers the data to Kafka. 
The last step consists on executing our workloads in stream mode. To do this, we have implemented an Extract, Transform and Load (ETL) program in order to process the received messages from Kafka. \textcolor{black}{The ETL routine consists on retrieving one tweet in its original format (JSON file), and  selecting a subset of attributes from the tweet such as hash-tag, text, geocoordinate, number of followers, name, surname and identifiers. All  the received messages are processed by our implemented workload. Then, they are stored using  ElasticSearch storage server, and possibly visualized with Kibana \cite{Kibana}.} 
Regarding the hardware configuration adopted in the Stream mode, we used one machine for Kafka and one machine for Zookeeper that allows the coordination between Kafka and Storm. For the processing task, the remaining machines are devoted to access the data in HDFS and to send it to Kafka server. 
 
\begin{figure*}[t]
\centering
\includegraphics[width=0.7\textwidth]{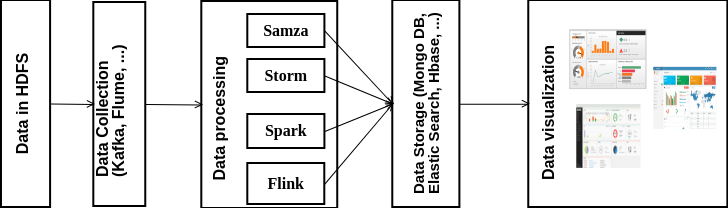}
\caption{\textit{Stream mode} scenario}
\label{fig:stream}
\end{figure*}

\end{itemize}

\begin{figure*}[t]
\centering
\includegraphics[scale=0.5]{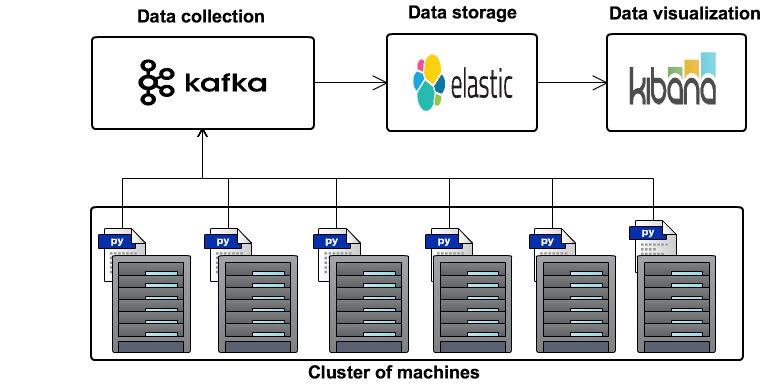}
\caption{Architecture of our personalized monitoring tool}
\label{fig:monitoring}
\end{figure*}

To allow monitoring resources usage according to the executed jobs, we have implemented a personalized monitoring tool as shown in Fig. \ref{fig:monitoring}. Our monitoring solution is based on three core components: (1) data collection module, (2) data storage module, and (3) data visualization module. To detect the states of the machines, we have implemented a Python script and we deployed it in every machine of the cluster. This script is responsible for collecting CPU, RAM, Disk I/O, and Bandwidth history. The collected data are stored in ElasticSearch, in order to be used in the evaluation step. The stored data are used by Kibana for monitoring and visualization purposes. \textcolor{black}{For our monitoring tests, we used a dataset of 50 GB of data for the WordCount workload, 10 millions examples for the K-means workload and the G5 dataset for the PageRank workload.} It is important to mention that existing monitoring tools like Ambari \cite{ambari} and Hue \cite{hue} are not suitable for our case, as they only offer real-time monitoring results. 

\subsection{Experimental results}
\subsubsection{Batch mode}
\textcolor{black}{In this section, we evaluate the scalability of the studied frameworks, and we measure their CPU, RAM, disk I/O usage, as well as  bandwidth consumption while processing. We also study the impact of several parameters and settings on the performance of the evaluated frameworks. }

\subsubsection*{Scalability}
\begin{figure}[t]
\centering
\includegraphics[scale=0.6]{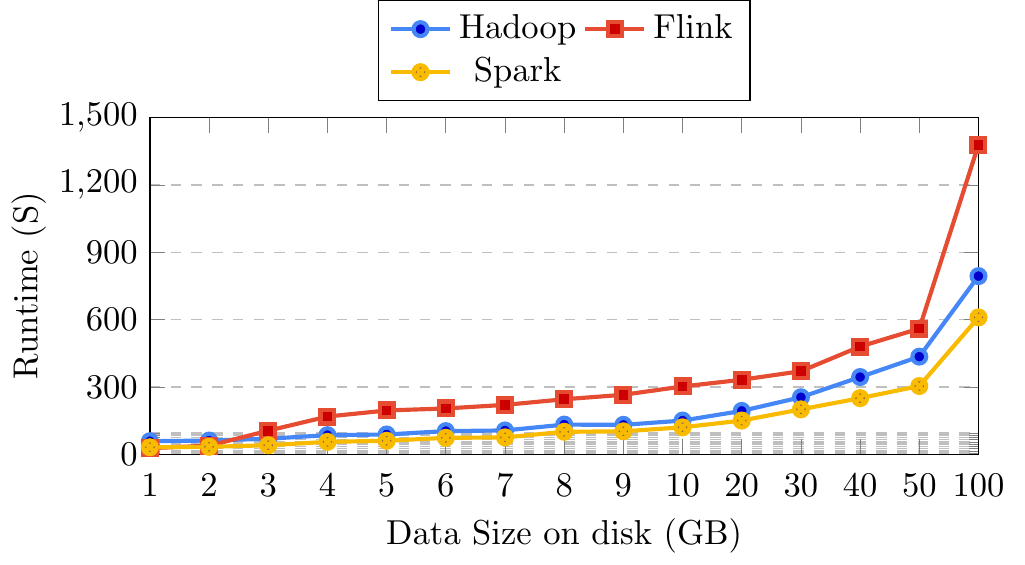}
\caption{Impact of the size of the data on the average processing time: case of big datasets}
\label{fig:bigdataset}
\end{figure}

\textcolor{black}{This experiment aims to evaluate the impact of the size of the data on the processing time. In this experiment, we used two simulations according to the size of data: (1) simulation with small datasets and (2) simulation with big datasets. }
\begin{figure}[t]
\centering
\includegraphics[scale=0.6]{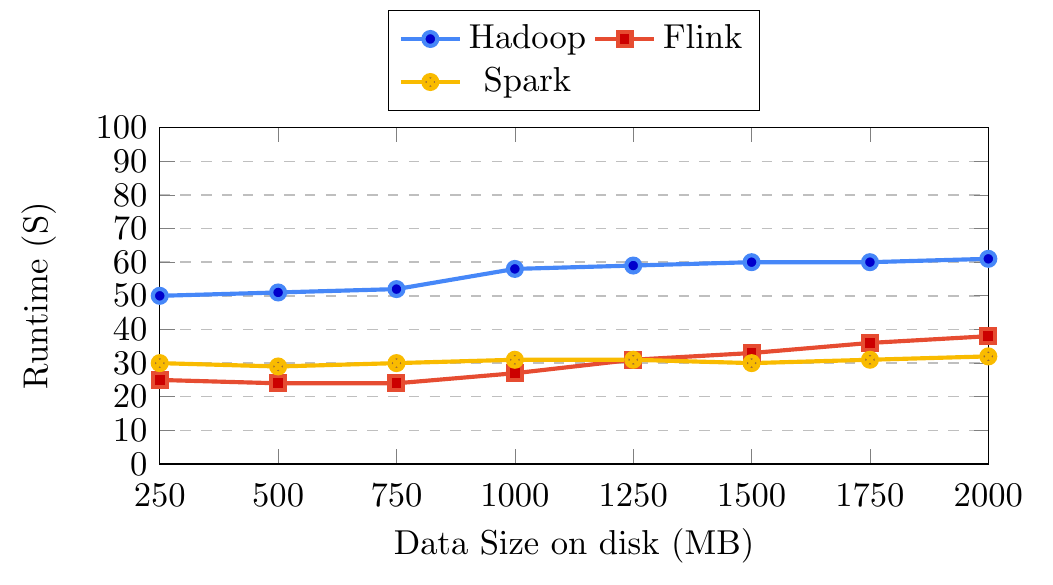}
\caption{Impact of the size of the data on the average processing time: case of small datasets}
\label{fig:smalldataset}
\end{figure}
\textcolor{black}{Experiments are conducted using the WordCount workload, with various datasets with a size on disk varying from 250 MB to 2 GB for the first simulation and from 1 GB to 100 GB for the second simulation.} Fig. \ref{fig:smalldataset} and Fig. \ref{fig:bigdataset} show the average processing time for each framework and for every dataset. As shown in Fig.  \ref{fig:smalldataset}, Spark is the fastest framework for all the datasets, Flink is the next and Hadoop is the lowest. Fig. \ref{fig:bigdataset} shows that Spark has kept its order in the case of big datasets and Hadoop showed good results compared to Flink. We also notice that Flink is faster than Hadoop only in the case of very small datasets.

Compared to Spark, Hadoop achieves data transfer by accessing the HDFS. Hence, the processing time of Hadoop is considerably affected by the high amount of Input/Output (I/O) operations. By avoiding I/O operations, Spark has gradually reduced the processing time.
It can also be observed that the computational time of Flink is longer than those of Spark and Hadoop in the case of big datasets. This is due to the fact that Flink sends its intermediate results directly to the network through channels between the workers, which makes the processing time very dependent on the cluster\textsc{\char13}s local network. In the case of small datasets, the data is transmitted quickly between workers. As shown in Fig. \ref{fig:smalldataset}, Flink is faster than Hadoop.
We also notice that Spark defines optimal time by using the memory to store the intermediate results as RDD objects.

\begin{figure}[t]
\centering
\includegraphics[scale=0.6]{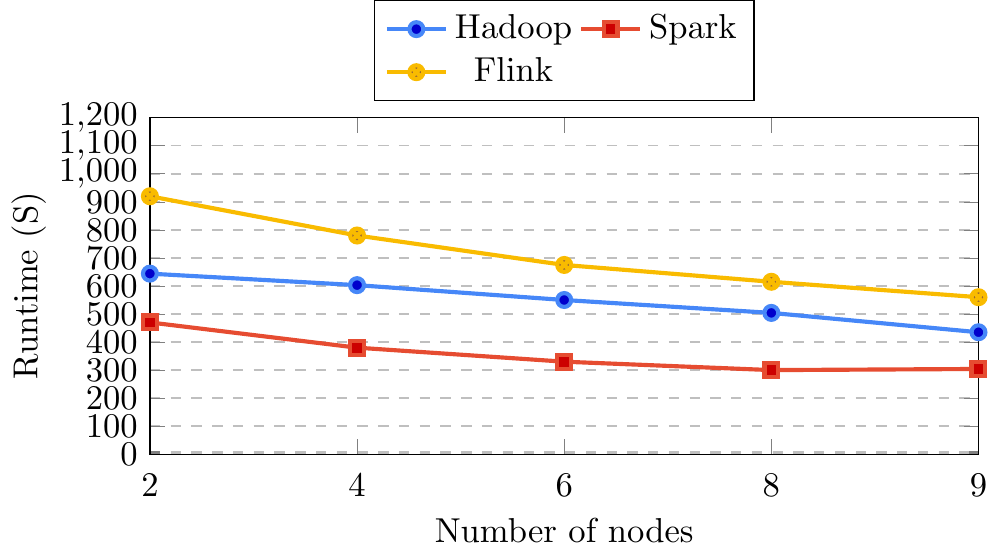}
\caption{Impact of the number of machines on the average processing time (WordCount workload with 50Gb of data)}
\label{fig:node}
\end{figure}

In the next experiment, we tried to evaluate the scalability and the processing time of the considered frameworks based on the size of the used cluster (the number of  machines in the cluster). Fig. \ref{fig:node} shows the impact of the number of the used machines on the processing time. Both Hadoop and Flink take higher time regardless of the cluster size, compared to Spark. Fig. \ref{fig:node} shows instability in the slope of Flink due to the network traffic. In fact, Flink jobs are modelled as a graph that is distributed on the cluster, where nodes represent Map and Reduce functions, whereas edges denote data flows between Map and Reduce functions. In this case, Flink performance depends on the network state that may affect intermediate results which are transferred from Map to Reduce functions across the cluster. Regarding Hadoop, it is clear that the processing time is proportional to the cluster size. In contrast to the reduced number of machines, the gap between Spark and Hadoop is reduced when the size of the cluster is large. This means that Hadoop performs well and can have close processing time in the case of a bigger cluster size. \textcolor{black}{The time spent by Spark is approximately between 450 seconds and 300 seconds for 2 to 6 nodes cluster. Furthermore, as the number of participating nodes increases, the processing time, yet, remains approximately equal to 290 seconds. This is explained by the processing logic of Spark. Indeed, Spark depends on the main memory (RAM) and the available resources in the cluster. In case of insufficient resources to process the intermediate results, Spark requires more RAM to store its intermediate results. This is the case of 6 to 9 nodes which explains the inability to improve the processing time even with an increased number of participating machines.}

\textcolor{black}{In the case of Hadoop, intermediate results are stored on disk. 
This explains the reduced execution time that reached 400 seconds in the case of 9 nodes, compared to 600 seconds when exploiting only 4 nodes.} To conclude, we mention that Flink allows to create a set of channels between workers, to transfer intermediate results between them. Flink does not perform Read/Write operations on disk or RAM, which allows accelerating the processing times, especially when the number of workers in the cluster increases. As for the other frameworks, the execution of jobs is influenced by the number of processors and the amount of Read/Write operations, on disk (case of Hadoop) and on RAM (case of Spark).
\subsubsection*{Iterative processing}
\begin{figure}[t]
\centering
\includegraphics[scale=0.6]{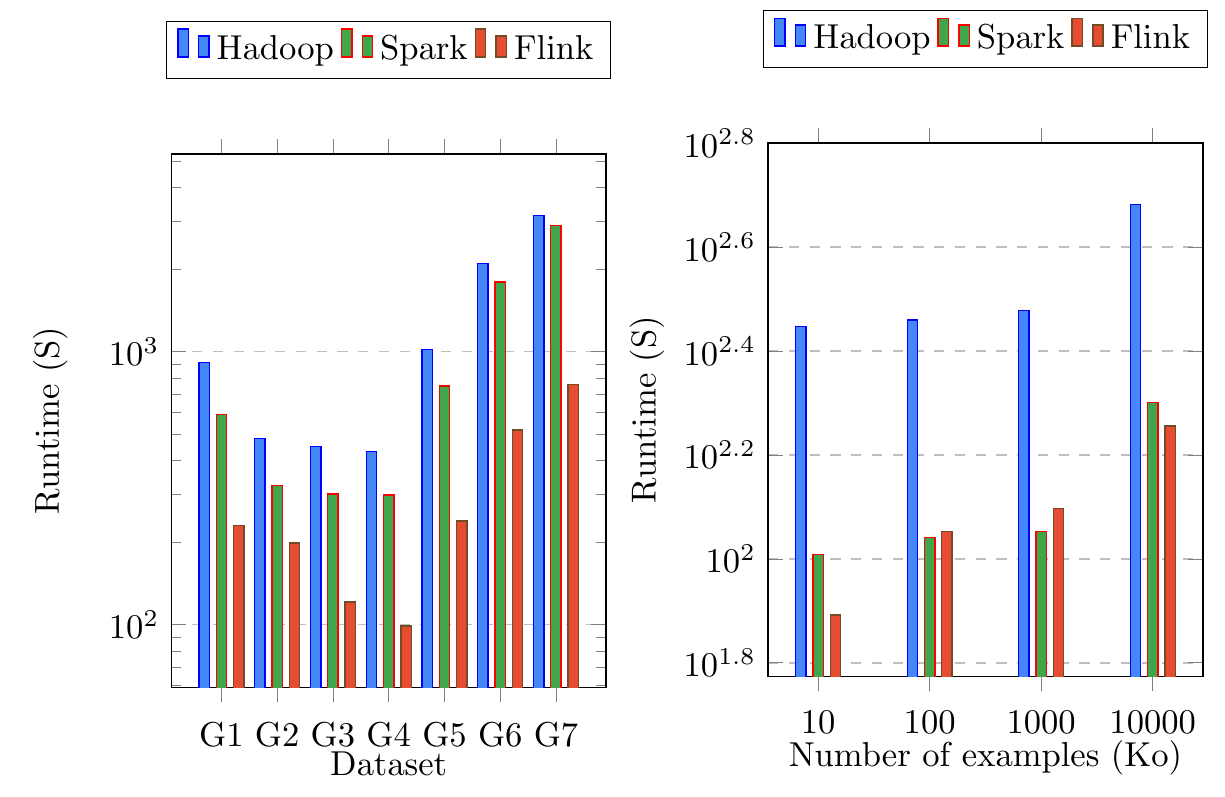}
\caption{Impact of iterative processing on the average processing time}
\label{fig:iterativeprocessing}
\end{figure}

In the next scenario, we tried to evaluate the studied frameworks in the case of iterative processing with both K-means and PageRank workloads. 
In Fig. \ref{fig:iterativeprocessing}, both use cases measure the impact of iterative computing on the studied frameworks. For K-means workload, we find that Spark and Flink are similar in  response time and they are faster compared to Hadoop. This can be explained by the fact that Hadooop writes the output results, for each iteration in the hard disk which makes Hadoop very slow. 
The PageRank workload is an iterative processing but it consumes more memory resources, which degrade performances in the case of Spark. 
In this case, Spark consumed all the available memory to create a new RDD object. Then, Spark applies its own strategy to replace the useless RDD and, when it does not fit in the memory, it slow down  the execution compared to both Flink and Hadoop. 

\begin{figure}[t]
\centering
\includegraphics[scale=0.6]{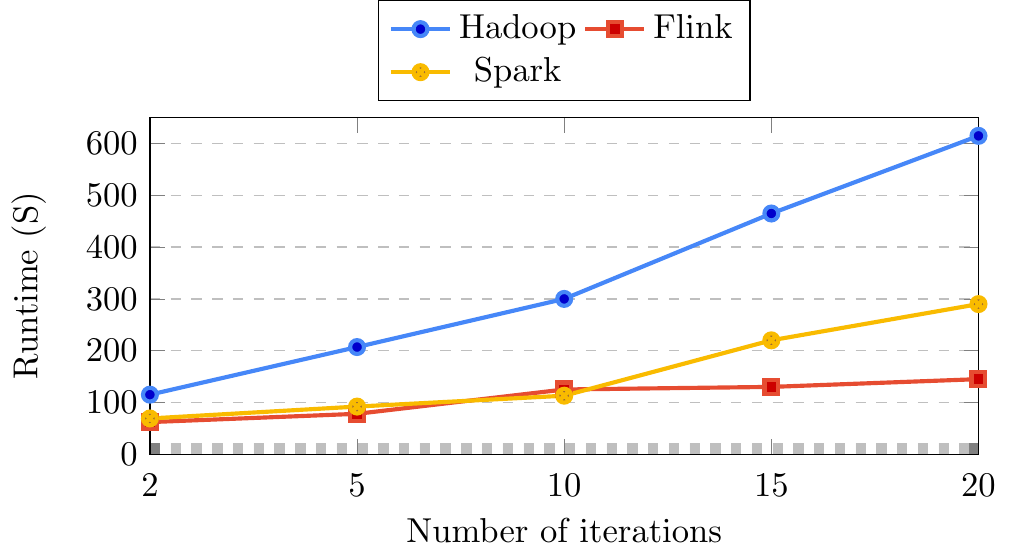}
\caption{Impact of the number of iterations on the average processing time (Kmeans workload with 10 million examples)}
\label{fig:iterations}
\end{figure}

\textcolor{black}{Through the next experiment, we try to show the impact of the number of iterations on the runtime. We tested our frameworks by running K-means on 10 millions examples in the training set. We varied the number of iteration in each simulation. }
As shown in Fig. \ref{fig:iterations}, with both Flink and Spark frameworks, the number of iterations has no significant influence on the execution time. One can conclude that the curve of Hadoop is characterized by an exponential slope whereas in the case of Spark and Flink the curves are characterized by a linear slope.
According to Fig. \ref{fig:iterations}, we can affirm that Hadoop is not the best choice for this kind of processing (iterative computing). 

\subsubsection*{Data partitioning}
In the next experiment, we try to show the impact of data partitioning on the studied frameworks. In our experimental setup, we used HDFS for storage. We varied the block size in our HDFS system and we run K-means with 10 iterations with all the used frameworks.
\begin{figure}[t]
\centering
\includegraphics[scale=0.6]{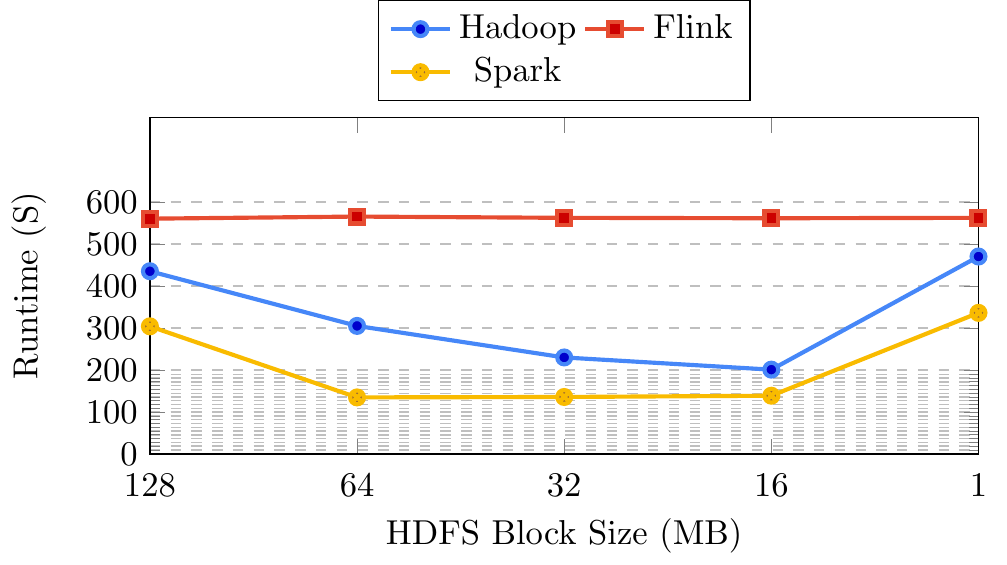}
\caption{Impact of HDFS block size on the runtime (Kmeans workload with 10 million examples and 10 iterations)}
\label{fig:blocksize}
\end{figure}
Fig. \ref{fig:blocksize} presents the  impact of the HDFS block size on the processing time. As shown in Fig. \ref{fig:blocksize}, the curves are inflated proportionally to the size of the HDFS block size for both Hadoop and Spark, while Flink does not imply any variation in the processing time. 
This can be explained by the degree of parallelism adopted by the studied frameworks. We mention that in Hadoop, the number of mappers is directly proportional to the input splits, which depends on HDFS block size. When we increase the number of splits, the degree of parallelism increases too. One possible solution to improve the processing time is to enhance the resource usage, but this is not always possible according to the Hadoop curve's behavior presented in Fig. \ref{fig:blocksize}. Note also that when we set a block of HDFS whose size is less than 16 MB, the processing time decreases as the number of input splits exceeds the number of cores in the cluster. 

\textcolor{black}{For Spark, we have almost the same results compared to Hadoop. Precisely, when Spark loads its data from HDFS, it converts or creates for each input split an RDD partition. In this case, the partition makes and provides the degree of parallelism, because Spark context program assigns for each worker an RDD partition.} 

As for Flink, each job is modeled as a directed graph, where nodes are reserved for data processing and edges represent data flow. In addition, each Flink job reserves a list of nodes in the graph to read the input data and to write the final results. In this case, the nodes responsible for reading input data, read from HDFS system and send the data as stream flow to other processing nodes. This mechanism makes Flink independent on the HDFS block size, as shown in Fig. \ref{fig:blocksize}.

\subsubsection*{Impact of the cluster manager}
\begin{figure*}[t]
\centering
\includegraphics[scale=0.6]{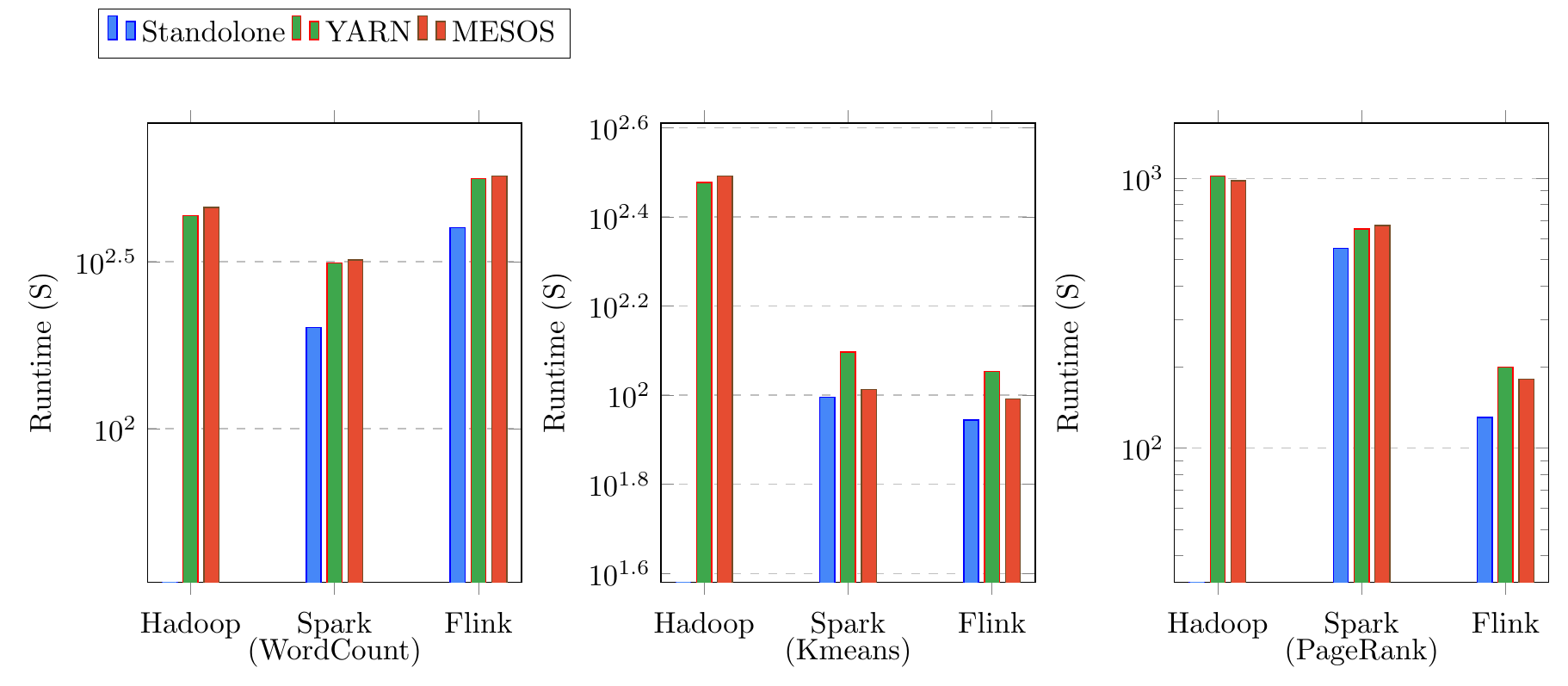}
\caption{\textcolor{black}{Impact of the cluster manager on the performance of the studied frameworks}}
\label{fig:clustermanager}
\end{figure*}

\textcolor{black}{
In our work, we mainly used YARN as a cluster manager. We also tried to evaluate the impact of the cluster manager on the performance of the studied frameworks. To do this, we compared Mesos, YARN and the standalone clusters manager of the studied frameworks. 
For our tests, we run the WordCount workload with 50 GB of data, K-means with 10 million examples and Pagerank with the G5 dataset (see Table \ref{tab:tab1}).
As shown in Fig. \ref{fig:clustermanager}, the standalone mode is faster than both YARN and MESOS. In fact, the standalone uses all the resources while executing a job, whereas both YARN and MESOS have a scheduler to run multiple jobs at once and share the cluster resources with all the submitted applications \cite{cm}.
}

\subsubsection*{Impact of bandwidth}
\textcolor{black}{
In order to study the impact of bandwidth consumption on the performance of the studied frameworks, we run the WordCount workload with 50 GB of data and we varied the bandwidth from 128 MB to 1GB.  
As shown in Fig. \ref{fig:bw}, Flink is bandwidth dependent. In fact, when the bandwidth increases, the response time decreases. This can be explained by the fact that each Flink job sends the data directly from a source to a calculating unit across the network. We also notice that Spark uses the network to, sometimes, migrate the data to the processing unit. Hadoop allows data locality, which means that Hadoop moves the computation close to where the actual data resides on the node.}

\begin{figure}[t]
\centering
\includegraphics[scale=0.6]{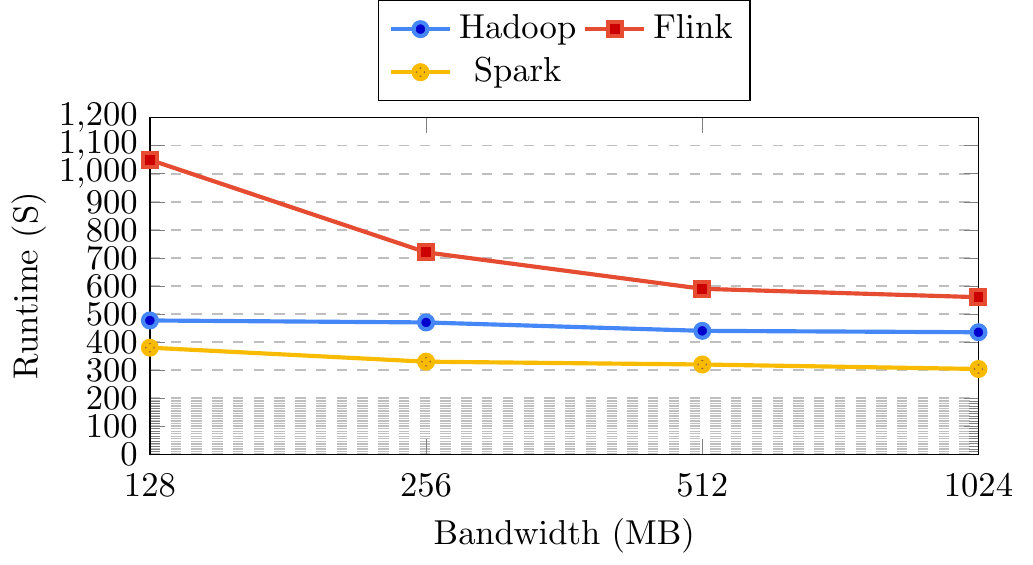}
\caption{\textcolor{black}{Impact of bandwidth on the performance of the studied frameworks}}
\label{fig:bw}
\end{figure}

\subsubsection*{Impact of some configuration parameters}
All the studied frameworks have a large list of configuration parameters, which can influence their behaviors. In order to understand the impact of the configuration parameters on the performance and the quality of the results, we try in this section to study some configuration parameters mainly related to the RAM and the number of threads in each framework.

In Hadoop, the Application Manager daemon distributes the Map and Reduce functions on the available slots of the cluster. 
To configure this aspect, we set both parameters   \textit{mapred.tasktracker.map.tasks.maximum} and \textit{mapred.tasktracker.reduce.tasks.maximum } in the \textit{site-mapred.xml} configuration file. These parameters represent respectively the maximum number of Map and Reduce tasks that will run simultaneously on a node. 
Note that Spark uses \textit{executor-cores} parameter and Flink uses \textit{slots} parameter to configure the number of executed threads in parallel. 

\begin{figure}[t]
\centering
\includegraphics[scale=0.6]{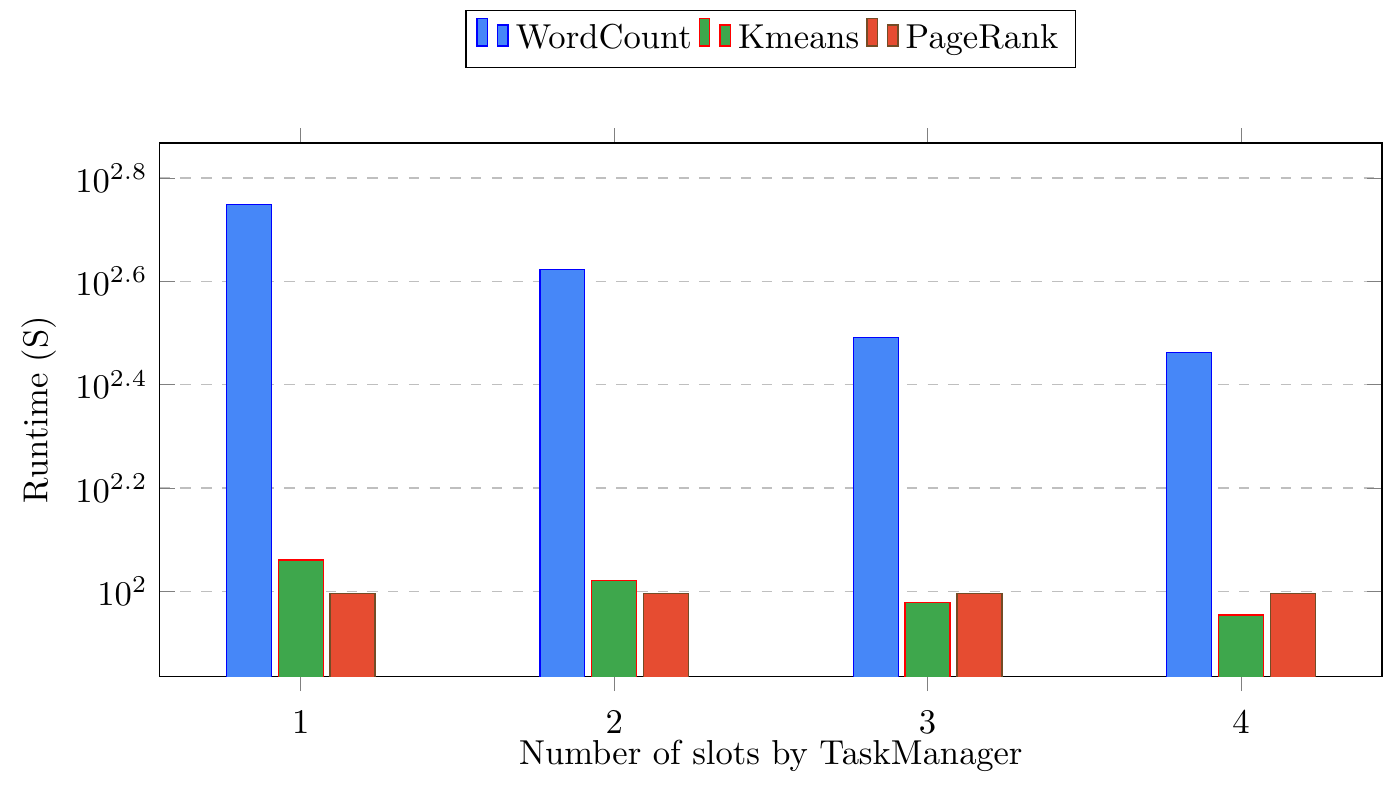}
\caption{Impact of parallelism parameters on the performance of Flink}
\label{fig:Flink_worker}
\end{figure}

\begin{figure}[t]
\centering
\includegraphics[scale=0.6]{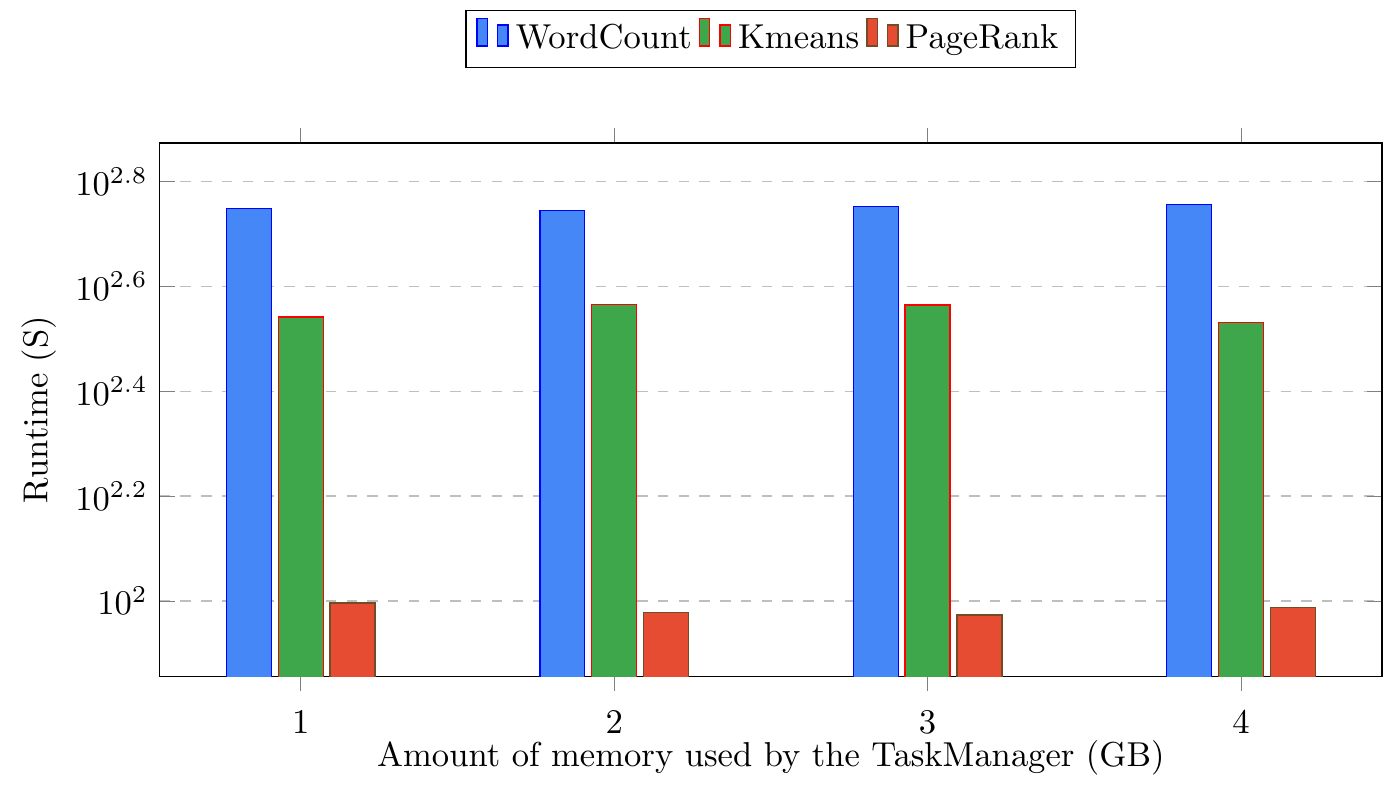}
\caption{Impact of the memory size on the performance of Flink}
\label{fig:Flink_ram}
\end{figure}

\begin{figure}[t]
\centering
\includegraphics[scale=0.6]{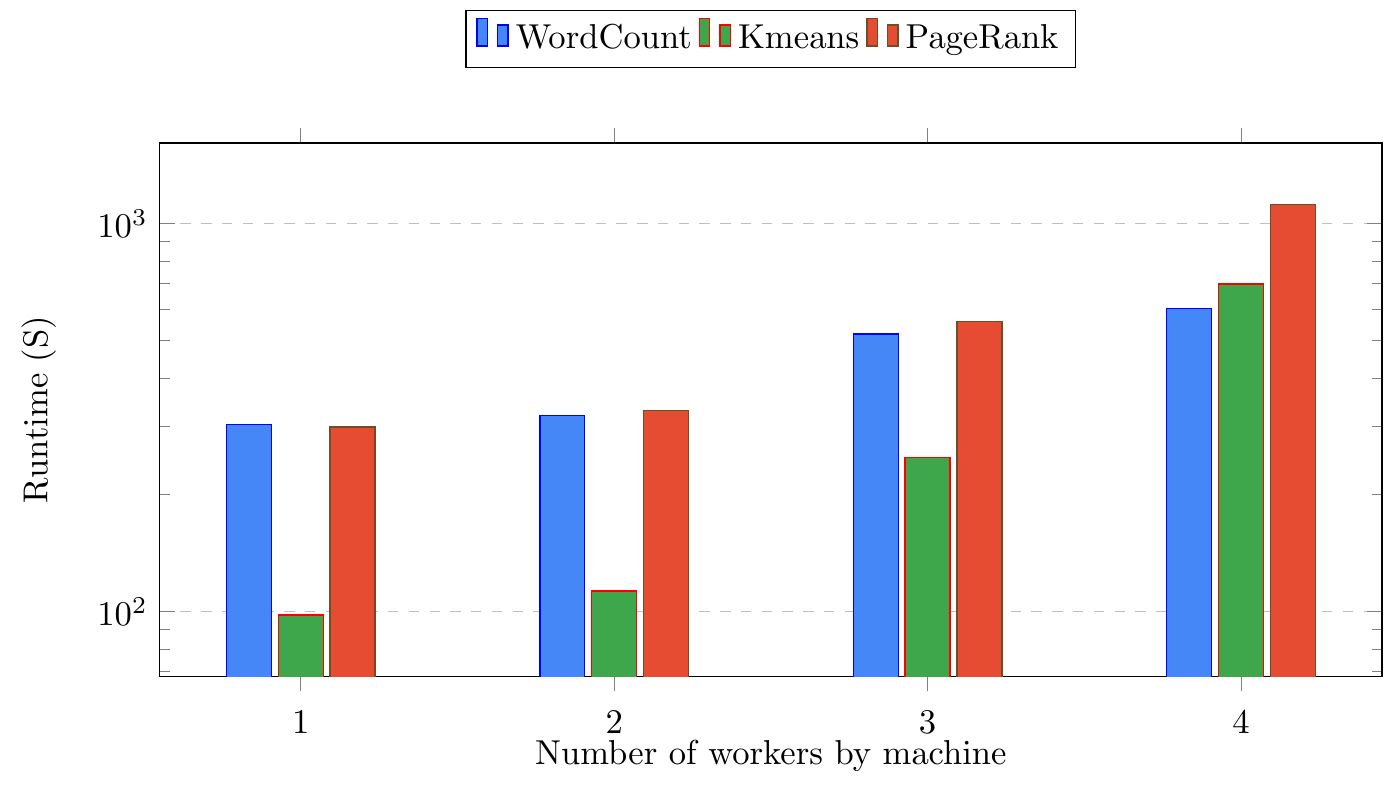}
\caption{Impact of the number of workers on the performance of Spark}
\label{fig:Spark_worker}
\end{figure}

\begin{figure}[t]
\centering
\includegraphics[scale=0.6]{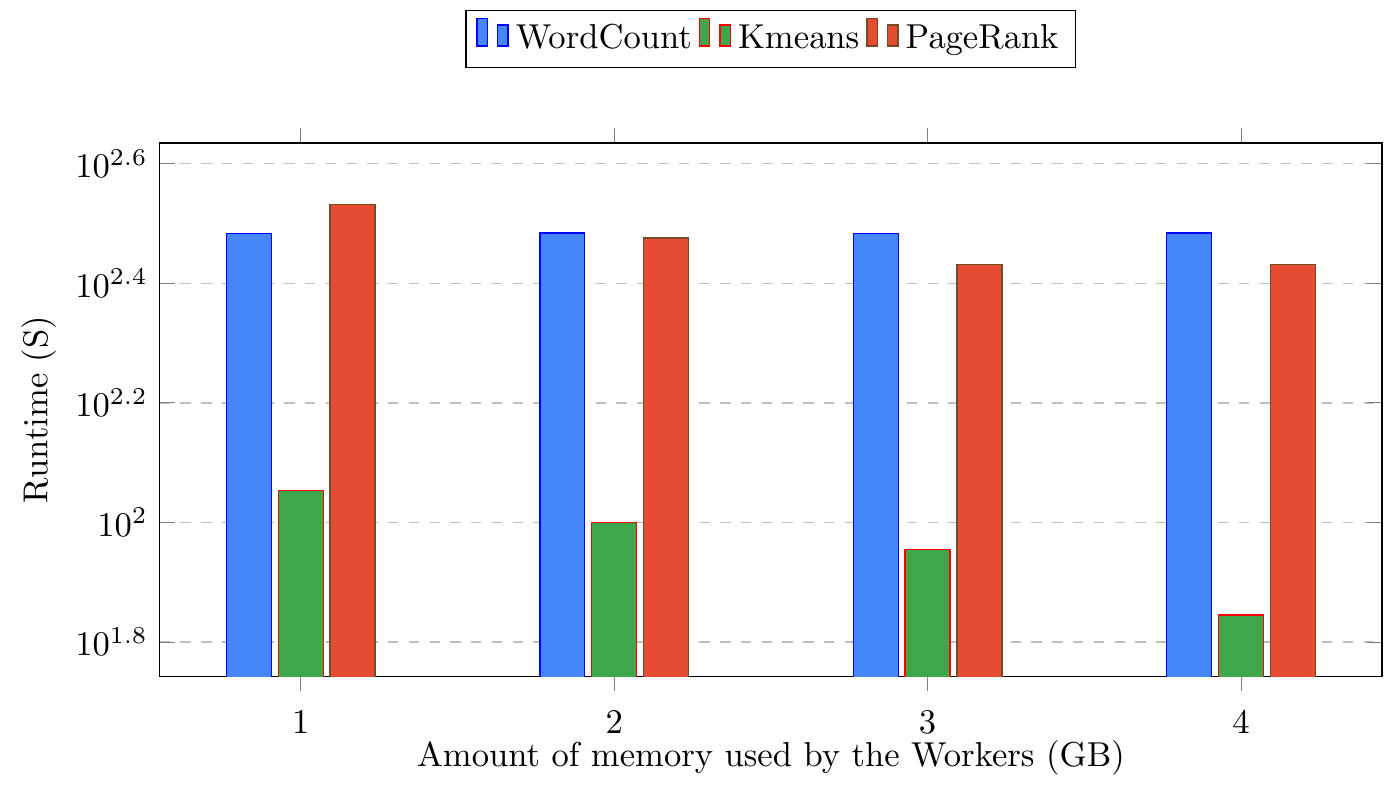}
\caption{Impact of the memory size on the performance of Spark}
\label{fig:Spark_ram}
\end{figure}

In order to define the amount of memory buffer, Hadoop uses the \textit{io.sort.mb} parameter in the  \textit{site-mapred.xml} configuration, Spark uses the \textit{executor-memory} parameter and Flink uses the \textit{taskManagerMemory} parameter.\\
\textbf{Flink configuration}.
In order to evaluate the impact of some configuration parameters on the performance of Flink, we first executed our workloads while varying the number of slots in each TaskManager. Then, we varied the amount of used memory by each TaskManager. Fig. \ref{fig:Flink_worker} presents the impact of the number of slots on the processing time of the WordCount workload. In fact, the latter is characterized by a high CPU resource consumption, that explains the reduction of the response time when the number of slots increases. Note that this is not the case with K-means and PageRank workloads. 
By analyzing Fig. \ref{fig:Flink_ram}, we notice that the memory resource does not have a large effect on the processing time in  these workloads, since Flink is based on sending the output results directly from one computing unit to another one without a high usage of disk or memory.
\\
\textbf{Spark Configuration}\\
The parallelism configuration in Spark requires the definition of the number of executor-cores by machine. In addition, the memory management is primordial as we must configure the memory for each worker. These two parameters are respectively \textit{executor-cores} and \textit{executor-memory}.
As shown in Fig. \ref{fig:Spark_worker}, when we increase the number of workers per machine the processing time increases too. This behavior may be related to the memory management of Spark. In fact, when the memory is shared and distributed on several slots, the slot of each worker will be limited which slows down the computing performance. In this case, it is advisable to limit the number of workers, if the machine has a limited  memory, and to maximize it proportionally to the capacity of the memory.
In the same context, we notice the importance of the memory size through Fig. \ref{fig:Spark_ram}. When we increase the memory the response time decreases. This behavior is not always valid because it depends on some other constraints such as the availability of other resources or traffic networks.
\\
\textbf{Hadoop Configuration}\\
To configure the number of slots on each node in a Hadoop cluster, we must set the two following parameters: (1) \textit{mapreduce.tasktracker.map.tasks.maximum} and (2) \textit{mapreduce.tasktracker.reduce.tasks.maximum}. These two parameters define the number of Map and Reduce functions that run simultaneously on each node of the cluster. These parameters maximize the CPU usage which can improve the processing time.

\begin{figure}[t]
\centering
\includegraphics[scale=0.6]{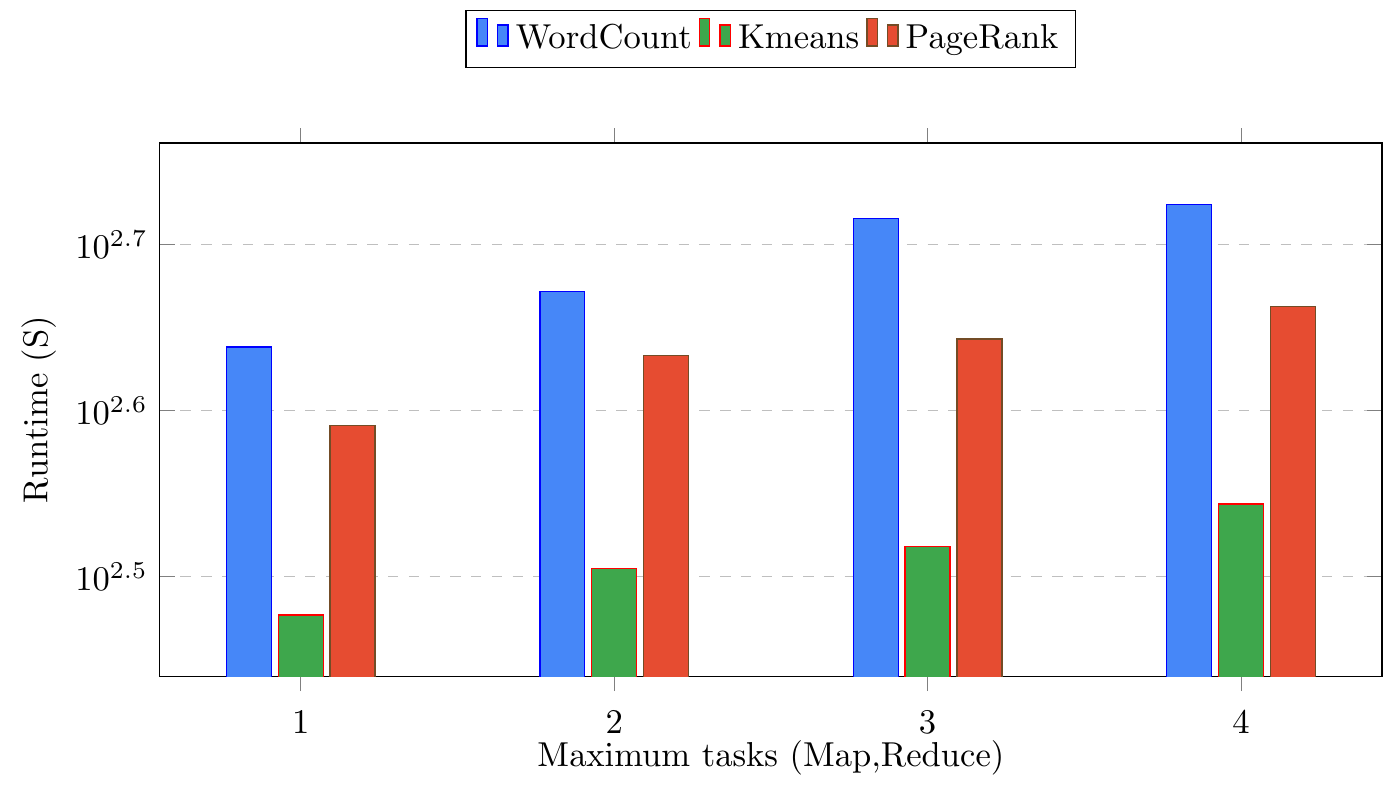}
\caption{Impact of the number of slots on the performance of Hadoop}
\label{fig:Hadoop_slot}
\end{figure}
Fig. \ref{fig:Hadoop_slot} shows the impact of the number of slots on the performance of Hadoop jobs. We find that the best performance is guaranteed when using two slots for both Map and Reduce functions. However, this value depends on the number of cores in each node of the cluster. In our case, we have four cores in each machine and the best value is two slots for Map and Reduce, since the other cores are reserved for both daemons DataNode and NodeManager. 
The same behavior is observed with the WordCount workload because this latter is based on CPU resource compared to the other workloads.
\begin{figure}[t]
\centering
\includegraphics[scale=0.6]{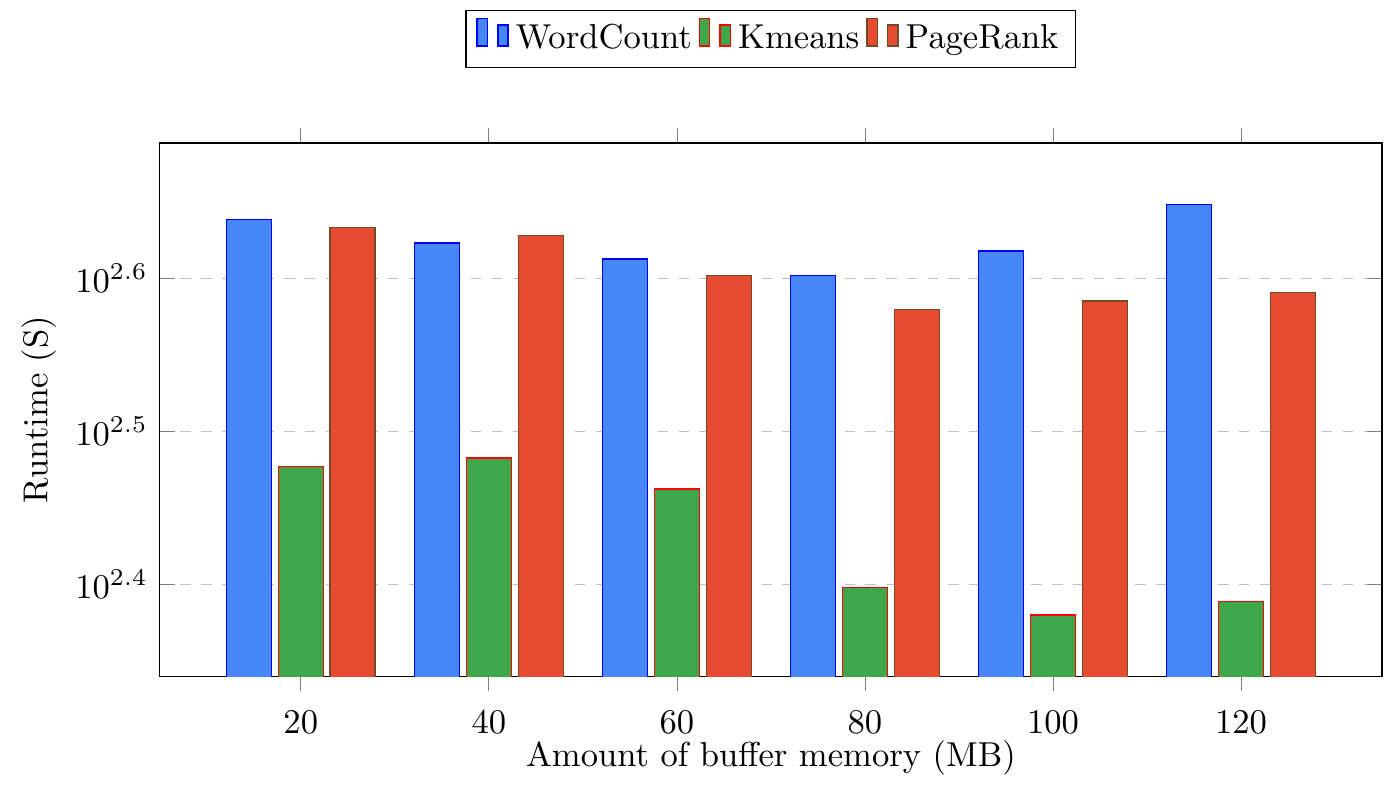}
\caption{Impact of the memory size on the performance of Hadoop}
\label{fig:Hadoop_ram}
\end{figure}
Among the characteristics of Hadoop, we note the use of the hard disk to write intermediate results between iterations or between Map and Reduce functions. Before writing data to the disk, Hadoop writes its intermediate data in a memory buffer. This memory can be configured through the \textit{io.sort.mb} parameter. In order to evaluate the impact of this parameter, we varied its values from 20 MB to 120 MB as illustrated in Fig. \ref{fig:Hadoop_ram}. \textcolor{black}{It is also clear that the processing time decreases subsequently and reaches 100 MB when we increase the value of \textit{io.sort.mb} parameter. A level of stability is achieved when the satisfaction of the computing units by this resource is guaranteed.}
\subsubsection*{Resources consumption}
\begin{figure*}[t]
\centering
\includegraphics[scale=0.6]{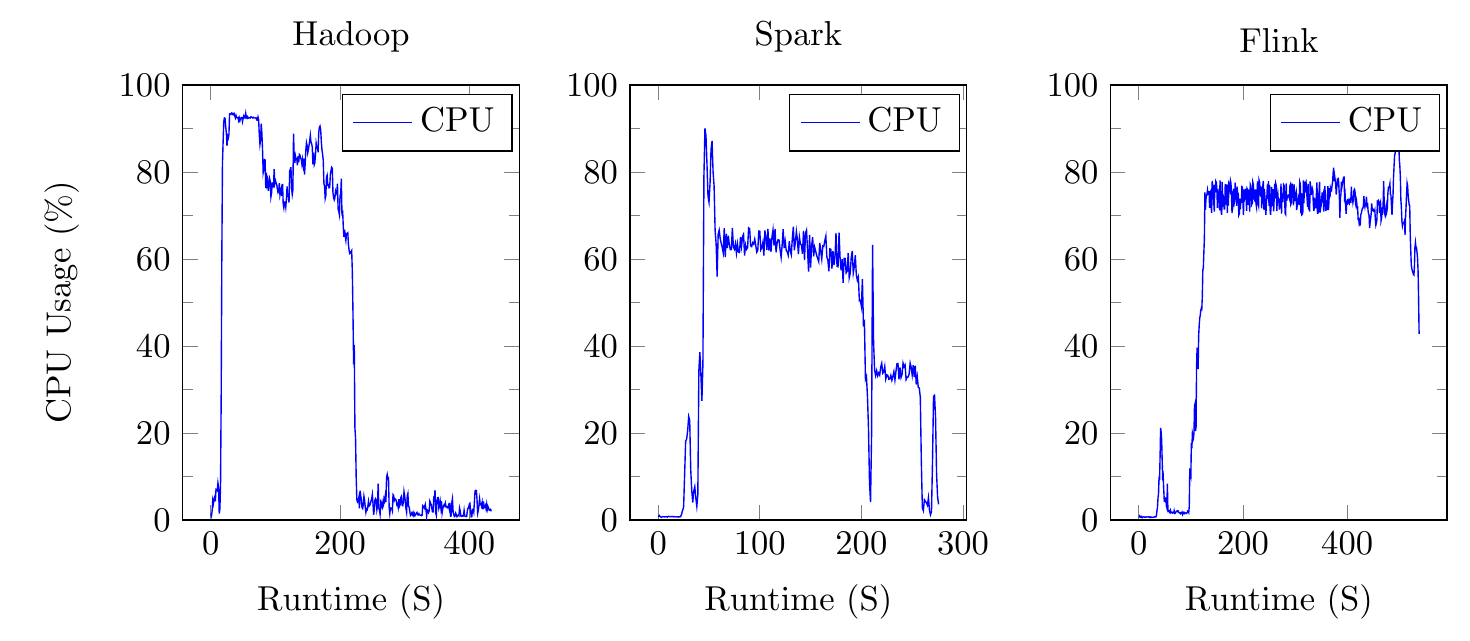}
\caption{CPU resource usage in \textit{Batch mode} scenario (WordCount workload with 50 GB of data)}
\label{fig:cpubatch}
\end{figure*}

\textbf{CPU consumption}\\
As shown in Fig. \ref{fig:cpubatch}, CPU consumption is approximately in direct ratio with the problem scale. However, the slopes of Flink (see Fig. \ref{fig:cpubatch}) are not larger than those of Spark and Hadoop because Flink partially exploits the disk and the memory resources, unlike Spark and Hadoop. Since Hadoop was initially modelled to frequently use the hard disk, the amount of Read/Write operations of the obtained processing results is high. Hence, Hadoop CPU consumption is important. In contrast, Spark mechanism relies on the  memory. This approach is not costly in terms of CPU consumption. 
Fig. \ref{fig:cpubatch} also shows Hadoop CPU usage. The processed data are loaded in the first 20 seconds. 
The next 220 seconds are devoted to execute the Map function. The reduce function is started in the last 180 seconds. The gaps in CPU usage (see Fig.\ref{fig:cpubatch}) is explained by the high number of Map functions (determined according to the data size and block size in HDFS), compared to the number of Reduce functions. When a Reduce function receives the totality of its key-value pairs (intermediate results generated by Map functions on the disk) assigned by the Application Manager, it starts the processing immediately. In the remaining processing time,  Hadoop writes the final results on disk through Reduce functions. As shown in Fig. \ref{fig:cpubatch}, the 
CPU usage is low because we have a single Reduce function (only one slot for writing the final results). 

\begin{figure}[t]
\centering
\includegraphics[scale=0.6]{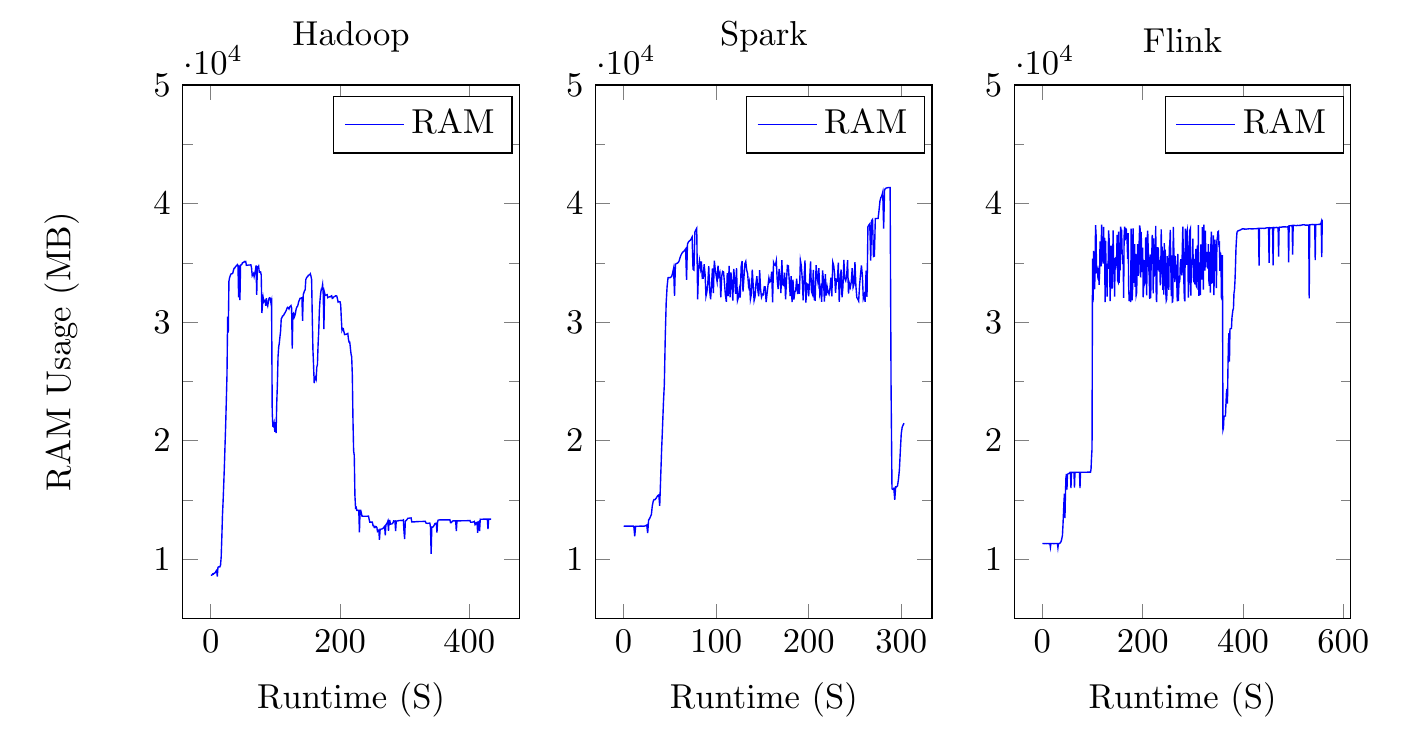}
\caption{RAM consumption in \textit{Batch mode} scenario (WordCount workload with 50 GB of data)}
\label{fig:rambatch}
\end{figure}

As for Spark CPU usage, Spark loads data from the disk to the memory during the first 20 seconds. The next 20 seconds are triggered to process the loaded data. In the second half time execution, each Reduce function processes its own data that come from the Map functions. In the last 10 seconds, Spark combines all the results of the Reduce functions. Although Flink execution time is higher than Spark and Hadoop, its overall processor consumption is high compared to Spark and Hadoop. Indeed, Flink divides the processing task into three steps. The first step is used to read the data from the source. The second step is used to process the data. The third step consists on writing the final results in the desired location. This behavior explains the use of the processor since the slots always listen to the data streams from the data sources. From Fig. \ref{fig:cpubatch}, it is clear that Flink maximizes the utilization of the CPU resource compared to Hadoop and Spark. 
\\
\textbf{RAM consumption}\\
Fig. \ref{fig:rambatch} plots the memory consumption of the studied frameworks. The RAM usage rate is almost the same for Flink and Spark especially in the Map phase. When the data fit into cache, Spark has a bandwidth between 35 GB and 40 GB. We notice that Spark logic depends mainly on RAM utilization. That is why, during 78.57\% of the total job execution time (120 seconds), the RAM is almost occupied (3.6 GB per node). Regarding Hadoop, only 45.83\% of the execution time (180 seconds) is used, and 35 GB of memory have been used during this time, where 2.6 GB in each node is reserved to the daemons of Hadoop. 

Another explanation of the fact that Spark RAM consumption is smaller than Hadoop is the data compression policies during data shuffle process. By default, Spark enabled the data compression during shuffle but Hadoop does not. Regarding Flink, RAM is occupied during 70\% of the total execution time (280 seconds), with 3.6 GB reserved for the daemons responsible for managing the cluster. The average memory usage is 40 GB. This is explained by the gradual triggering of Reduce functions after receiving the intermediate results that came from Map functions. Flink models each task as a graph, where its constituting nodes represent a specific function and edges denote the flow of data between those nodes. Intermediate results are sent directly from Map to groupBy functions without massively using RAM and disk resources.
\\
\textbf{Disk I/O usage}
\begin{figure}[t]
\centering
\includegraphics[scale=0.5]{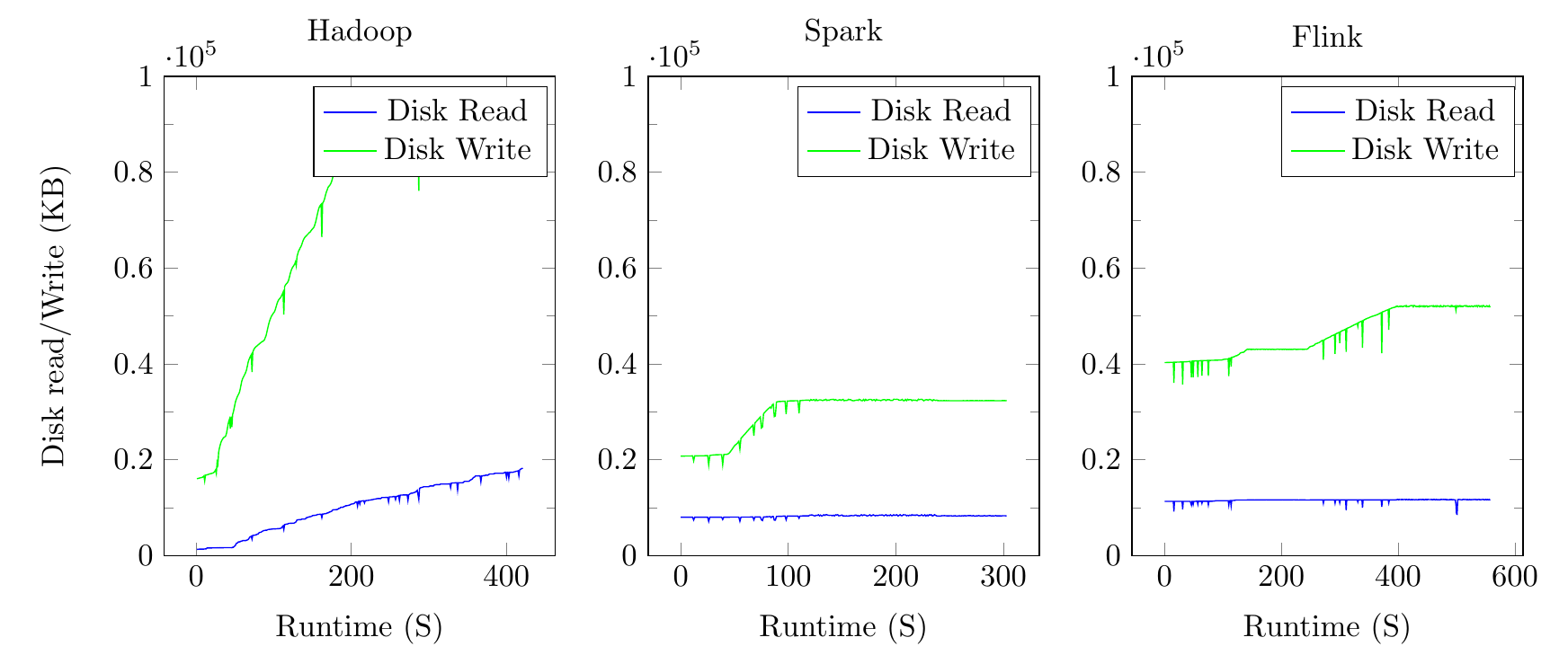}
\caption{Disk I/O usage in \textit{Batch mode} scenario (WordCount workload with 50 GB of data)}
\label{fig:discbatch}
\end{figure}

\begin{figure}[t]
\centering
\includegraphics[scale=0.5]{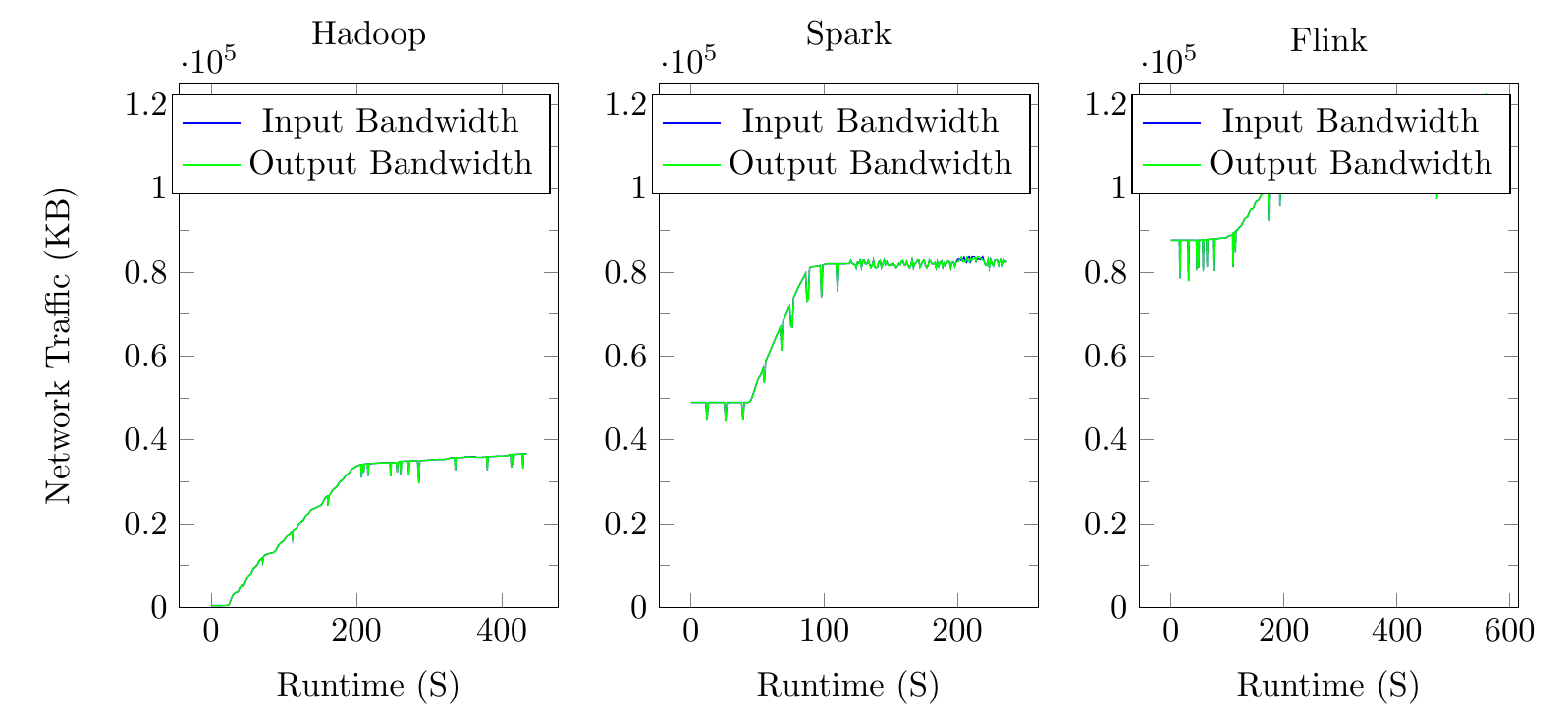}
\caption{Bandwidth resource usage in \textit{Batch mode} scenario (WordCount workload with 50 GB of data)}
\label{fig:bwbatch}
\end{figure}

Fig. \ref{fig:discbatch} shows the amount of disk usage by the studied frameworks. We find that Hadoop frequently accesses the disk, as the amount of write operations is about 90 MB/s. This is not the case for Spark (about 30 MB/s) as this framework is memory-oriented. As for Flink which shares a similar behavior with Spark, the disk usage is very low compared to Hadoop (about 50 MB/s). Indeed, Flink uploads, first, the required processing data to the memory and, then, distributes them among the candidate workers.
\\
\textbf{Bandwidth resource usage}\\
As shown in Fig. \ref{fig:bwbatch}, Hadoop has a best traffic utilization. As for Flink, it surpasses both Spark and Hadoop in traffic utilization. The amount of data exchanged per second is high compared to Spark and Hadoop (37 MB/s for Hadoop, 120 MB/s for Flink, and 70 MB/s for Spark). The massive use of bandwidth resources could be attributed to the streaming of data in Flink jobs between data source and data processing nodes.

As mentioned before, Flink sends directly the outputs of the Map functions to the next  functions through channels of data transmission between these functions, which explains the very high utilization of bandwidth. As for Spark, the data compression policies during the data shuffle process allow this framework to record intermediate data in temporal files and compress them before their submission from one node to another. This has a positive impact on bandwidth resource usage as the intermediate data are transferred in a reduced size. Regarding Hadoop, the data placement strategy helps the latter to optimize the use of bandwidth resource. It is also important to mention that, when a MapReduce job is submitted to the cluster, the resource manager assigns the Map functions to the nodes of the cluster while minimizing the data exchange between the nodes. 

\subsubsection{Stream mode}
In stream experiments, we measure CPU, RAM, disk I/O usage and bandwidth consumption of the studied frameworks while processing tweets, as described in  Section \ref{sec:experimProtocol}.

\begin{figure}[t]
\centering
\includegraphics[scale=0.6]{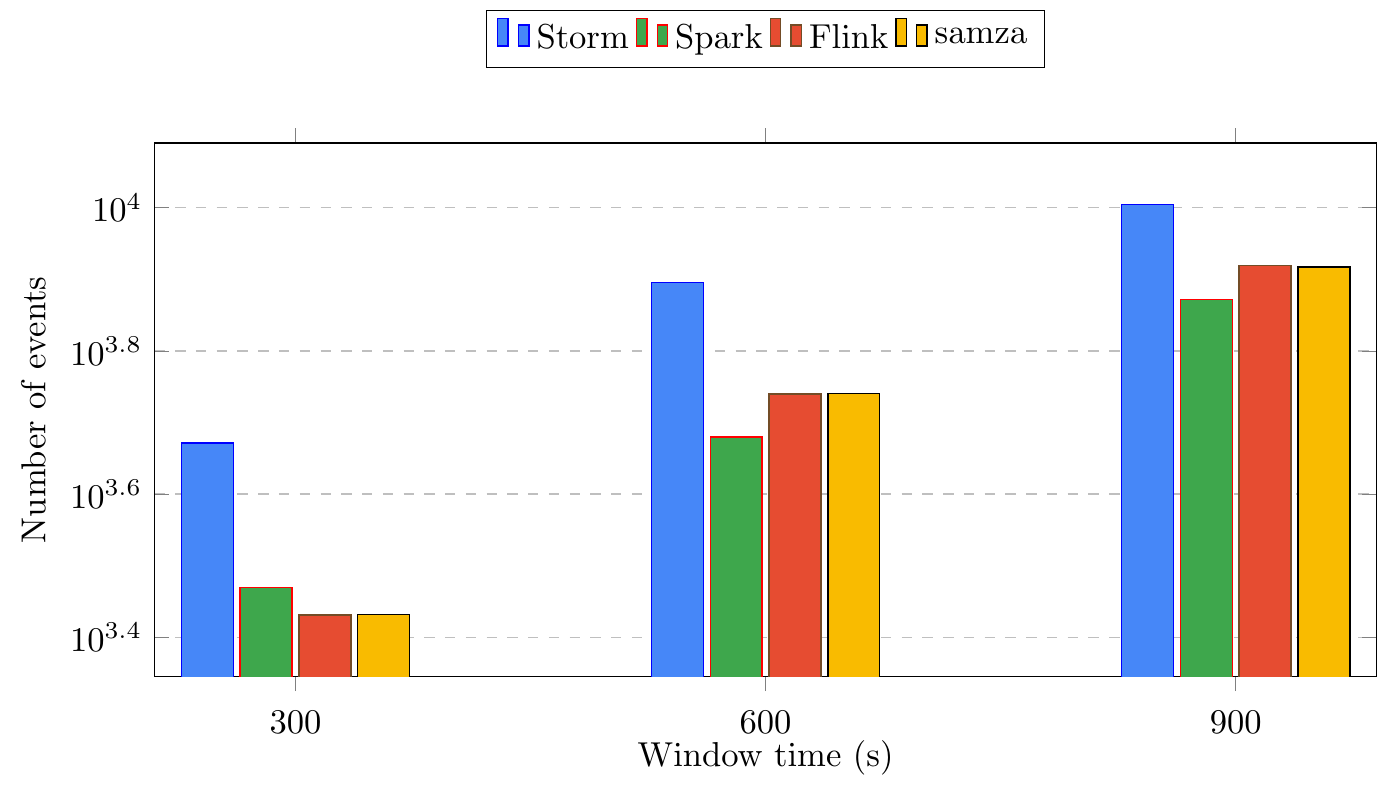}
\caption{Impact of the window time on the number of processed events (100 KB per message)}
\label{fig:window1}
\end{figure}

The goal here is to compare the performance of the studied frameworks according to the number of processed messages within a period of time. In the first experiment, we send a tweet of 100 KB (in average) per message. Fig. \ref{fig:window1} shows that Flink, Samza and Storm have better processing rates compared to Spark. This can be explained by the fact that the studied frameworks use different values of window time. The values of window time of Flink, Samza  and Storm are much smaller than that of Spark (milliseconds vs seconds).

\begin{figure}[t]
\centering
\includegraphics[scale=0.6]{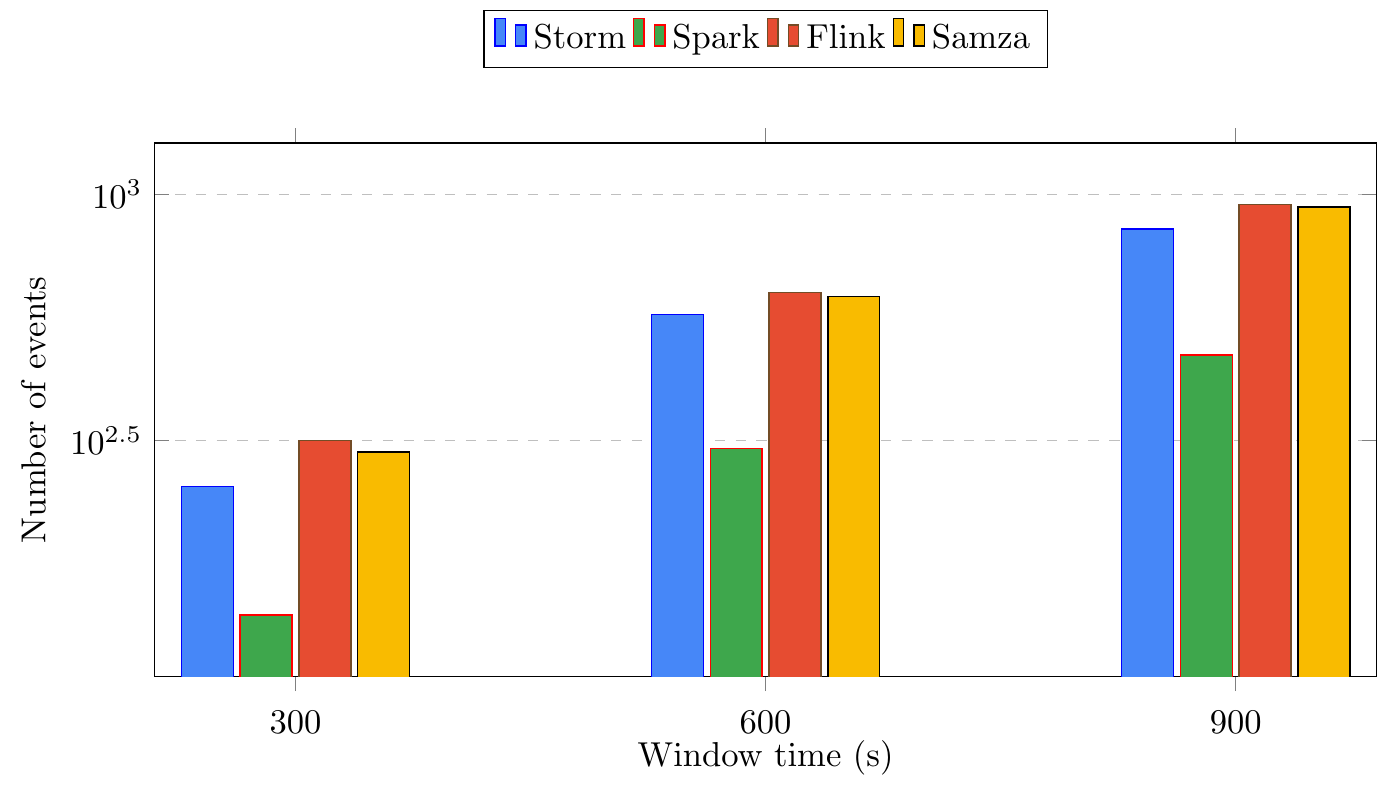}
\caption{Impact of the window time on the number of processed events (500 KB per message)}
\label{fig:window2}
\end{figure}

\begin{figure*}[t]
\centering
\includegraphics[scale=0.6]{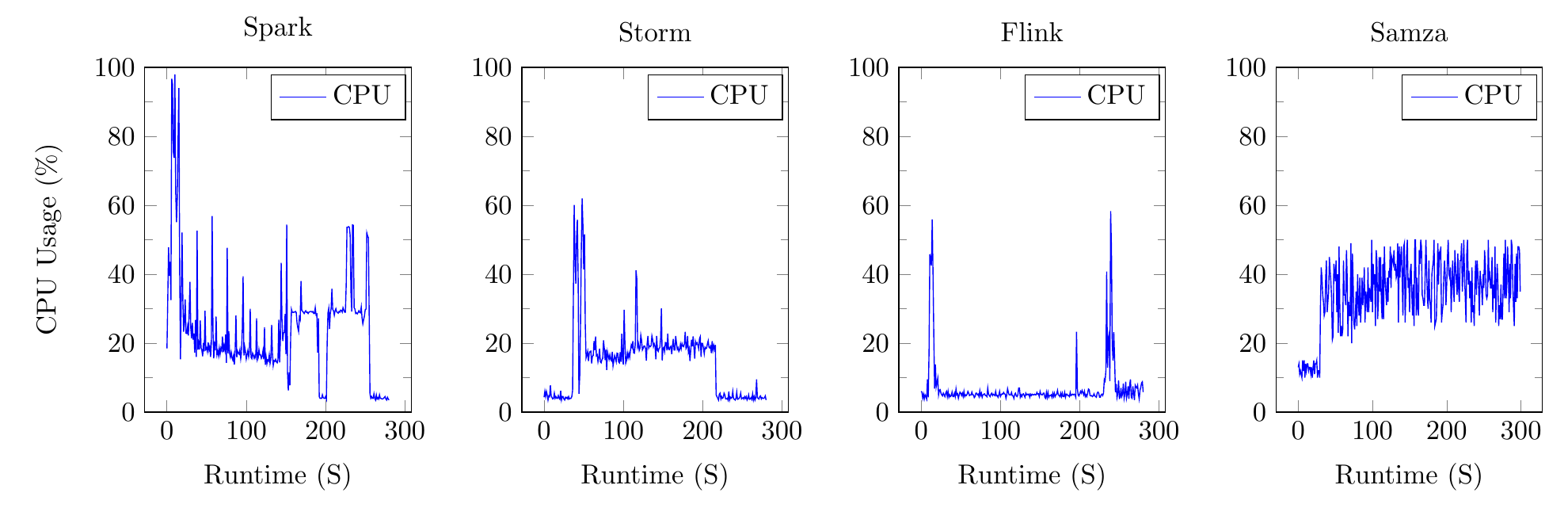}
\caption{CPU consumption in \textit{Stream mode} scenario (with 100 KB per message use case)}
\label{fig:cpu}
\end{figure*}

In the next experiment, we changed the sizes of the processed messages. We used 5 tweets per message (around 500 KB per message). The results presented in Fig. \ref{fig:window2} show that Samza and Flink are very efficient compared to Spark, especially for large messages. \\
\textbf{CPU consumption}\\
\textcolor{black}{Results have shown that the number of events processed by Storm (10085) is close to that processed by Flink (8320) despite the larger-size nature of events in Flink compared to Samza and Storm.}
In fact, the window time\textsc{\char13}s configuration of Storm allows to rapidly deal with the incoming messages. 
Fig.~\ref{fig:cpu} plots the CPU consumption rate of Flink, Storm and Spark. 

\begin{figure*}[t]
\centering
\includegraphics[scale=0.5]{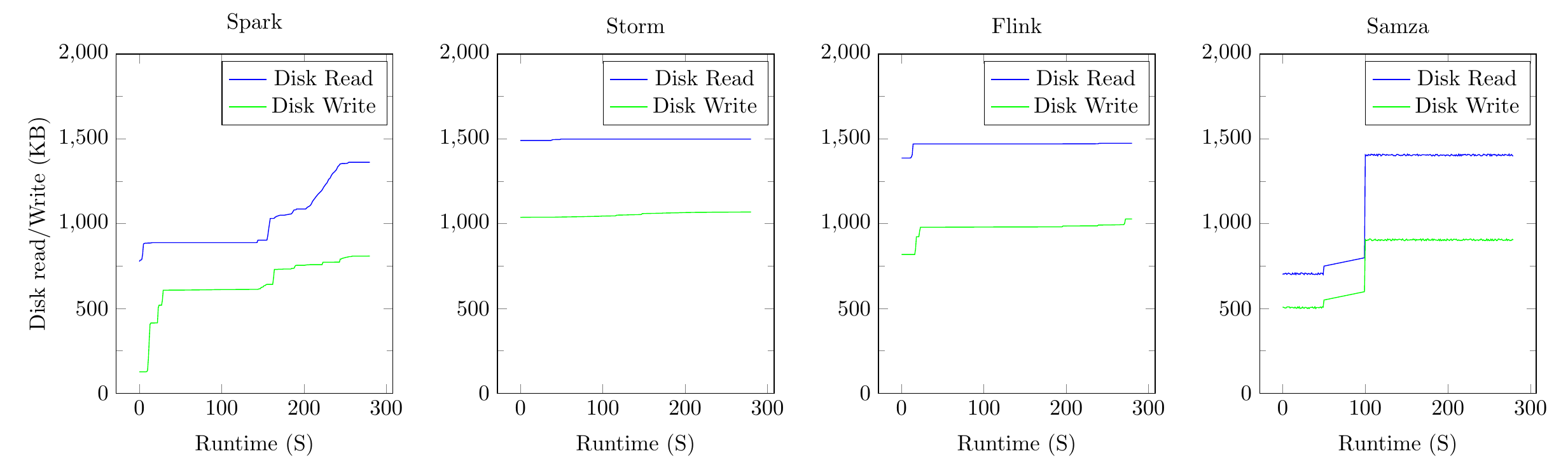}
\caption{Disk usage in \textit{Stream mode} scenario (with 100 KB per message)}
\label{fig:disc}
\end{figure*}

As shown in Fig.~\ref{fig:cpu}, Flink CPU consumption is low compared to Spark, Samza and Storm. Flink exploits about 10\% of the available CPU to process 8320 events, whereas Storm CPU usage varies between 15\% and 18\% when processing 10085 events. However, Flink may provide better results than Storm when CPU resources are more exploited. In the literature, Flink is designed to process large messages, unlike Storm which is only able to deal with small messages (e.g., messages coming from sensors). Unlike Flink, Samza and Storm, Spark collects events' data every second and performs processing task after that. Hence, more than one message is processed, which explains the high CPU usage of Spark. Because of Flink's pipeline nature, each message is associated to a thread and consumed at each window time. Consequently, this low volume of processed data does not affect the CPU resource usage.
\textcolor{black}{
Samza exploits about 55\% of the available CPU because it is based on the concept of  virtual cores and, each job or partition is assigned to a number of virtual cores. So, we can deploy several threads (one for each partition), which explains the intensive CPU usage compared to the other frameworks.
}
\\
\textbf{RAM consumption} \\
\begin{figure*}[t]
\centering
\includegraphics[scale=0.6]{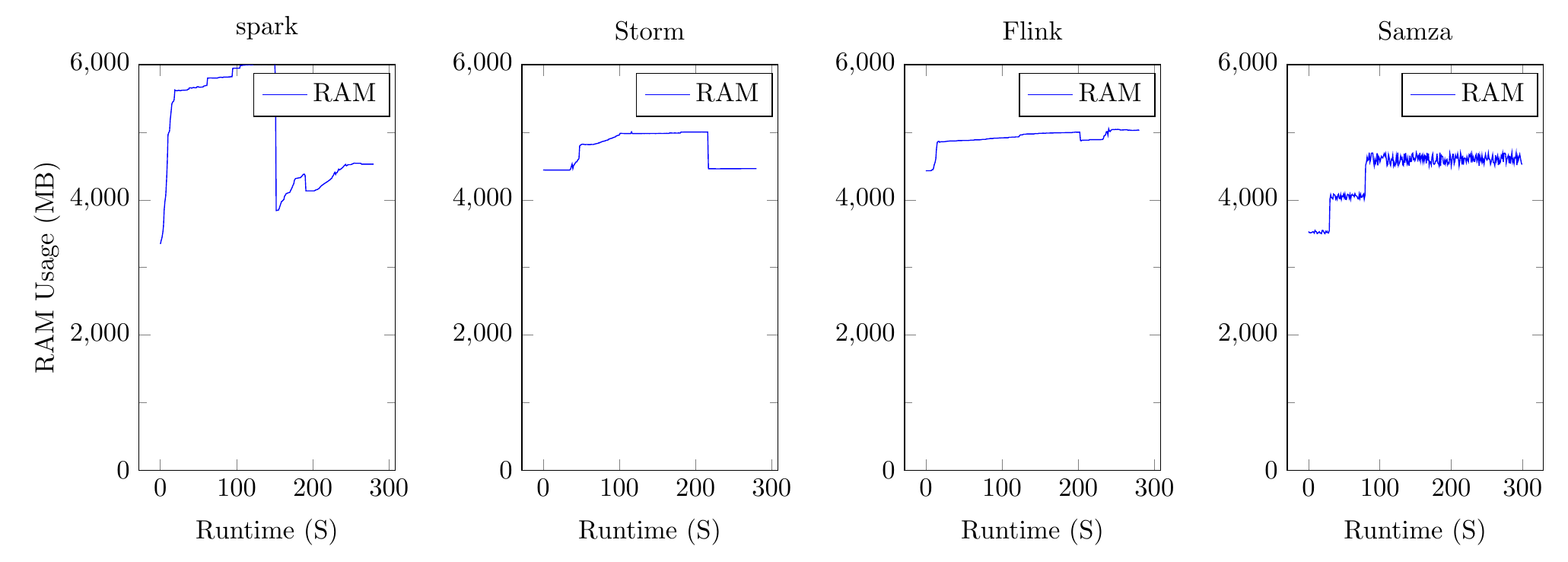}
\caption{RAM consumption in \textit{Stream mode} scenario (with 100 KB per message)}
\label{fig:ram}
\end{figure*}
Fig. \ref{fig:ram} shows the cost of event stream processing in terms of RAM consumption. Spark reached 6 GB (75\% of the available resources) due to its in-memory behavior and its ability to perform in micro-batch (process a group of messages at a time). Flink, Samza and Storm did not exceed 5 GB (around 61\% of the available RAM) as their stream mode behavior consists in processing only single messages. Regarding Spark, the number of processed messages is small. 
Hence, the communication frequency with the cluster manager is low. In contrast, the number of processed events is high for Flink, Samza and Storm, which explains the important communication frequency between the frameworks and their Daemons (i.e. between Storm and Zookeeper, or between Flink and Yarn). Indeed, the communication topology in Flink is predefined, whereas the communication topology in the case of Storm is dynamic because Nimbus (the master component of Storm) searches periodically the available nodes to perform processing tasks.
\\
\textbf{Disk R/W usage}\\
Fig. \ref{fig:disc} depicts the amount of disk usage by the studied frameworks. The curves denote the amount of Read/Write operations. The amounts of Write operations in Flink and Storm are almost close.
Flink, Samza and Storm frequently access the disk and are faster than Spark in terms of the number of processed messages. As discussed in the above sections, Spark framework is an in-memory framework which explains its lower disk usage.
\\
\textbf{Bandwidth resource usage}\\
\textcolor{black}{ As shown in Fig. \ref{fig:bw}, the amount of data exchanged per second varies between 375 KB/s and 385 KB/s in the case of Flink, and varies between 387 KB/s and 390 KB/s in the case of Storm and about 400 Mb/s in the case of Samza. This amount is high compared to Spark as its bandwidth usage did not exceed 220 KB/s. This is due to the reduced frequency of serialization and migration operations between the cluster nodes, as Spark processes a group of messages at each operation. Consequently, the amount of exchanged data is reduced, while  Storm, Samza and Flink are designed for the stream processing. }
\begin{figure*}[t]
\centering
\includegraphics[scale=0.5]{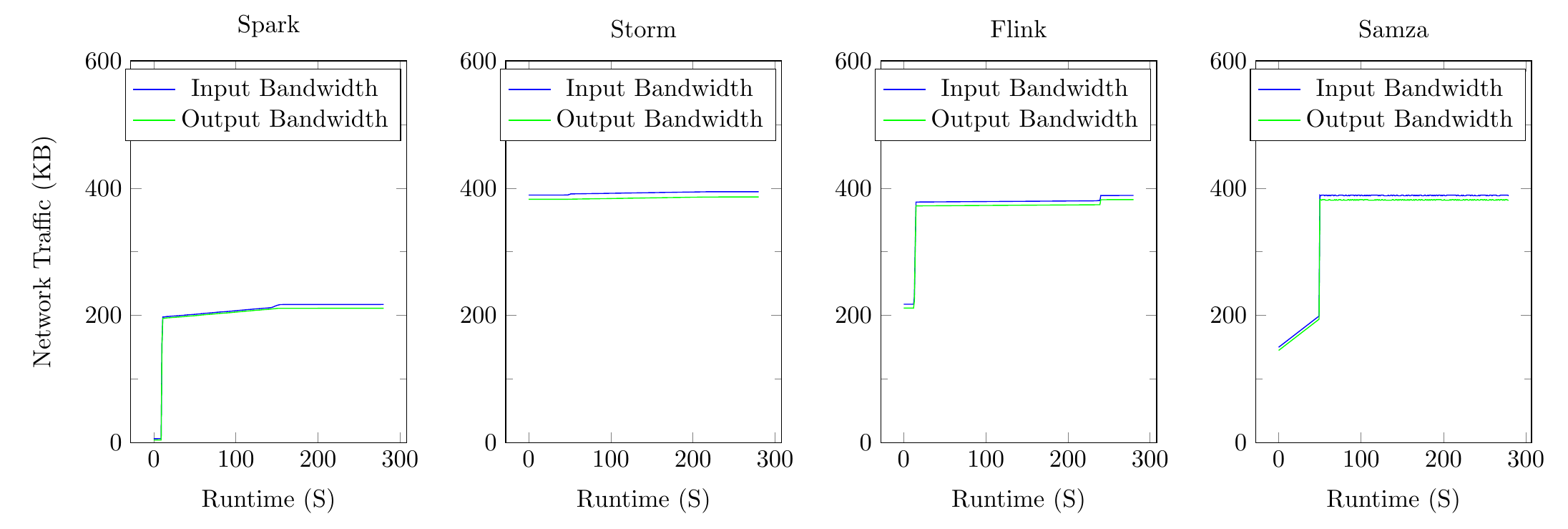}
\caption{Traffic bandwidth in \textit{Stream mode} scenario (with 100 KB per message use case)}
\label{fig:bw}
\end{figure*}

\subsection{Summary of the evaluation}
From the above presented experiments, it is clear that Spark can deal with large datasets better than Hadoop and Flink. Although Spark is known to be the fastest framework due to the concept of RDD, it is not a suitable choice in the case of intensive memory processing tasks. Indeed, intensive memory applications are characterized by the massive use of memory (creation of RDD objects at each transformation operation). This process degrades the performance of Spark since the SparkContext will be led to find the unused RDD and remove them in order to get more free memory space.
The carried experiments in this work also indicate that Hadoop performs well on the whole. However, it has some limitations regarding the writing of intermediate results in the hard disk and requires a considerable processing time when the size of data increases, especially in the case of iterative applications.
According to the resource consumption results in batch mode, we can conclude that Flink maximizes the use of CPU resources compared to both frameworks Spark and Hadoop. This good exploitation is relative to the pipeline  technique of Flink which minimizes the period of idle resources. However, it is characterized by high demands on network resource compared to Hadoop. In fact, this resource consumption explains why Flink is faster than Hadoop. 
\textcolor{black}{
In the stream scenario, Flink, Samza and Storm are quite similar in terms of data processing. In fact, they are originally designed for stream processing. We also notice that Flink is characterized by its low latency, since it is based on pipe-lined processing and on message passing processing technique, whereas Spark is based on Java Virtual Machine (JVM) and belongs to the category of batch mode frameworks.
Each Samza job is divided into one or more partitions and each partition is processed in an independently container or executor, which shows best results with large stream messages.
Another important aspect to be considered while tuning the used framework is the cluster manager. In the standalone mode, the resource allocation in Spark and Flink are specified by the user during the submission of its jobs whereas using a cluster manager such as Mesos or YARN, the allocation of the resources is done automatically.}

\section{Best Practices}
\label{sec:best}
In the previous section, two major processing approaches (batch and stream) were studied and compared in terms of speed and resource usage. Choosing the right processing model is a challenging problem, given the growing number of frameworks with similar and various services \cite{sakr16}. This section aims to shed light on the strengths of the above discussed frameworks when exploited in specific fields including stream processing, batch processing, machine learning and graph processing.
\subsection{Stream processing}
As the world becomes more connected and influenced by mobile devices and sensors, stream computing emerged as a basic capability of real-time applications in several domains, including monitoring systems, smart cities, financial markets and manufacturing \cite{Bajaber2016}. However, this flood of data that comes from various sources at high speed always needs to be processed in a short time interval. In this case, Storm and Flink may be considered, as they allow pure stream processing.
The design of in-stream applications needs to take into account the frequency and the size of incoming events data. In the case of stream processing, Apache Storm is well-known to be the best choice for the big/high stream oriented applications (billions of events per second/core). As shown in the conducted experiments, Storm performs well and allows resource saving, even if the stream of events becomes important. 
\subsection{Micro-batch processing}
In case of batch processing, Spark may be a suitable framework to deal with periodic processing tasks such as Web usage mining, fraud detection, etc. In some situations, there is a need for a programming model that combines both batch and stream behaviour over the huge volume/frequency of data in a lambda architecture. In this architecture, periodic analysis tasks are performed in a larger window time. Such behaviour is called micro-batch. For instance, data produced by healthcare and IoT applications often require combining batch and stream processing. In this case frameworks like Flink and Spark may be good candidates \cite{Landset2015}. Spark micro-batch behaviour allows to process datasets in larger window times. Spark consists of a set of tools, such as SparkMLLIB and Spark Stream that provide rich analysis functionalities in micro-batch. Such behaviour requires regrouping the processed data periodically, before performing analysis task.
\subsection{Machine learning algorithms}
\textcolor{black}{
Machine learning algorithms are iterative in nature \cite{Landset2015}. They are widely used to process huge amounts of data and to exploit the opportunities hidden in big data \cite{Zhou2017ml}.  
Most of the above discussed frameworks support machine learning capabilities through a set of libraries and APIs. FlinkML library includes implementations of k-Means clustering algorithm, logistic regression, and Alternating Least Squares (ALS) for recommendation \cite{Chakrabarti:2008:DMK:1521583}. Spark has more efficient set of machine learning algorithms such as Spark MLlib \cite{8258338} and MLI \cite{mli}. Spark MLlib is a scalable and fast library that is suitable for general needs and most areas of machine learning. Regarding Hadoop framework, Apache Mahout aims to build scalable and performant machine learning applications on top of Hadoop. 
}
\subsection{Big graph processing}
The field of large graph processing has attracted considerable attention because of its huge number of applications, such as the analysis of social networks  \cite{Giatsidis:2011:ECC:2055438.2055765}, Web graphs \cite{webgraphs} and bioinformatics \cite{bigbioinfo} \cite{mrsimlab}. It is important to mention that Hadoop is not the optimal programming model for graph processing \cite{alberto3}. This can be explained by the fact that Hadoop uses coarse-grained tasks to do its work, which are too heavyweight for graph processing and iterative algorithms \cite{Landset2015}. In addition, Hadoop can not cache intermediate data in memory for faster performance. We also notice that most of Big Data frameworks provide graph-related libraries (e.g., Graphx \cite{Xin:2013:GRD:2484425.2484427} with Spark and Flinkgelly \cite{flink} with Flink). Moreover, many graph processing systems have been proposed~\cite{bdr2016}. Such frameworks include Pregel~\cite{15malewicz2010pregel}, Graphlab~\cite{graphlab}, Bladyg~\cite{bladyg} and Trinity~\cite{trinity}. 

\section{Conclusions}
 \label{sec:conc}
In this work, we surveyed popular frameworks for large-scale data processing. 
After a brief description of the main paradigms related to Big Data problems, 
we presented an overview of the Big Data frameworks Hadoop, Spark, Storm and Flink. 
We presented a categorization of these frameworks according to some main features such as the used programming model, the type of data sources, the supported programming languages and whether the framework allows iterative processing or not. We also conducted an extensive comparative study of the above presented frameworks on a cluster of machines and we highlighted best practices while using the studied Big Data frameworks. 



\bibliographystyle{abbrv}
\bibliography{biblio}  

\end{document}